\documentclass[iop,numberedappendix]{emulateapj}
\usepackage[dvips]{color}

\usepackage[]{graphicx}
\usepackage[active]{srcltx}
\usepackage{aas_macros}
\usepackage{amsmath}

\begin{document}

\title{The imaging properties of the Gas Pixel Detector as a focal plane polarimeter}

\author{S. Fabiani, E. Costa, E. Del Monte, F. Muleri, P. Soffitta, A. Rubini}
\affil{INAF-IAPS, via del Fosso del Cavaliere 100, 00133 Roma (Italy)}


\author{R. Bellazzini, A. Brez, L. de Ruvo, M. Minuti, M. Pinchera, C. Sgr\'o, G. Spandre}
\affil{INFN Sezione di Pisa, Largo B. Pontecorvo, 3 - 56127 Pisa (Italy) }


\author{D. Spiga, G. Tagliaferri, G. Pareschi, S. Basso, O. Citterio}
\affil{INAF-Osservatorio Astronomico di Brera, via Brera 28, 20121 Milano (Italy)}


\author{V. Burwitz, W. Burkert, B. Menz, G. Hartner}
\affil{Max-Planck-Institut f\"ur extraterrestrische Physik, Gautinger Str. 45, 82061 Neuired (Germany)}

\email{sergio.fabiani@iaps.inaf.it}

\begin{abstract}

X-rays are particularly suited to probe the physics of extreme objects. However, despite the enormous improvements of X-ray Astronomy in imaging, spectroscopy and timing, polarimetry remains largely unexplored. 
We propose the photoelectric polarimeter Gas Pixel Detector (GPD) as an instrument candidate to fill the gap of more than thirty years of lack of measurements. 
The GPD, in the focus of a telescope, will increase the sensitivity of orders of magnitude. Moreover, since it can measure the energy, the position, the arrival time and the polarization angle of every single photon, allows to perform polarimetry of subsets of data  singled out from the spectrum, the light curve or the image of source.
The GPD has an intrinsic very fine imaging capability and in this work we report on the calibration campaign carried out in 2012 at the PANTER X-ray test facility of the Max-Planck-Institut f\"ur extraterrestrische Physik of Garching (Germany) in which, for the first time, we coupled it to a JET-X optics module with a focal length of 3.5~m and an angular resolution of 18~arcsec at 4.5~keV.
This configuration was proposed in 2012 aboard the X-ray Imaging Polarimetry Explorer (XIPE) in response to the ESA call for a small mission. 
We derived the imaging and polarimetric performance for extended sources like Pulsar Wind Nebulae and Supernova Remnants as case studies for the XIPE configuration, discussing also possible improvements by coupling the detector with advanced optics, having finer angular resolution and larger effective area, to study with more details extended objects.

\end{abstract}

\keywords{X-ray polarimetry, X-ray telescope, angular resolution}

\section{Introduction}
X-ray Astronomy obtained important results  so far by using imaging, spectroscopy and timing. New observational techniques are required to refine theoretical models and remove degeneracies by adding new observational parameters.
X-ray polarimetry would allow for introducing the degree and the angle of polarization that relate closely to the emission mechanism and to the source geometry. However, despite the enormous improvements in X-ray Astronomy, polarimetry remained largely unexplored. 
The first detection of polarized X-rays from an astrophysical source was obtained for the \object[M1]{Crab nebula} in the 1971, by means of a sounding rocket experiment \citep{Novick1972}. The result was later confirmed and the polarization was precisely measured with a degree of ($19.2\% \pm  1.0\%$) at 2.6 keV and ($19.5\% \pm  2.8\%$) at 5.2 keV \citep{Weisskopf1976, Weisskopf1978} by the polarimeter on board the OSO-8 satellite. This result ultimately proved the synchrotron origin of the X-ray emission of the nebula and remains, still today, the only precise non-zero one since the `70s, while upper limits were measured by \cite{Hughes1984}. 
The need to measure the polarization of high energy emission of many other sources remains urgent, and new technological solutions are available today. 

So far only few measurements of non-imaging polarimetry, with moreover a low significance, where performed. 
Non-imaging polarimetry averages the polarization of all subsystems  within the field of view. For extended sources this can result in a substantial spoiling of the physical information. The reduction of measured polarization arises from the cancellation of the polarization vectors coming from regions with a different polarization state. This is crucial for extended sources such as the PWNe and the SNRs.
X-ray polarimeters with imaging capability would allow to overcome this problem and to obtain polarization maps of extended sources. Moreover, imaging is useful to increase the signal to noise ratio for polarimetry  by developing analysis strategies aimed to reduce the contamination of the emission due to source regions or emission components different from the ones of interest (see in particular the study of the pulsar signal in PWNe of Sect.~\ref{sec:PWN}).
This improvement is possible only with a detector having simultaneously both polarimetric and imaging capabilities. A combination of an imager detector and a non-imaging polarimeter would not be adequate to this aim.

The Gas Pixel Detector (GPD) \citep{Costa2001, Bellazzini2003} exploits the photoelectric effect to perform polarimetry and it is able also to make simultaneously spectral and timing measurements. The tracks of photoelectrons are produced in gas with a charge content proportional to the photon energy. From their initial emission direction the polarimetric measurement is derived, while the image is obtained as a map of the photoionization locations.
The GPD is the most advanced 2-D imaging polarimeter with high polarimetric sensitivity and spatial resolution with respect to other instruments. 
For example CCDs were considered to perform polarimetry \citep{Tsunemi1992, Buschhorn1994} by exploiting the border effect among neighbour pixels to detect photoelectron polarization. However, this technique is heavily limited by systematics due to the small range of photoelectrons in silicon (only $\simeq$1.5$\mu$m at 10 keV) with respect to the pixel size. An other technique by \cite{Sakurai2004} exploits CCDs to detect the UV scintillation images of photoelectron tracks in a Capillary Gas Proportional Counter. Photoelectric effect in gas is exploited also in TPCs for GEMS \citep{Black2010}. A high quantum efficiency is obtained at expense of imaging, while its  1-D imaging capability is very much blurred by inclined penetration (see Sect.~\ref{sec:GPD}), due to focusing, in astronomical implementation  \citep{Jahoda2010} with a consequent much larger background. At higher energies the Compton scattering polarimeter by \cite{Hayashida2012} has some imaging capability, with an angular resolution of few arcmin. In this case the spatial resolution depends basically on the  width (few millimetres) of the scattering rods.

The intrinsic imaging capability of the GPD was already studied by \cite{Soffitta2013a} who measured the \textit{spatial resolution} of the detector alone (with a narrow parallel X-ray beam).
In our work we study the performance of a GPD  combined with an X-ray telescope and compare them with predictions. 
From simulation studies \citep{Fabiani2009, Lazzarotto2010} we expect that the GPD, if coupled with an X-ray optical module with an intrinsic angular response in the range of a fraction of arcminute, should allow for imaging without a significant loss of performance with respect to the intrinsic angular resolution of the telescope.
In this work we report about this, by proving it experimentally for the first time. Even if this paper is focused on the analysis of the imaging properties, we discuss, briefly, also the relationship between polarization and grazing incidence reflection. This is useful to clarify what is the expected limit of spurious polarization induced by optics and why we have no concern about the feasibility of polarimetry by means of the GPD coupled with X-ray telescopes.
The GPD was placed at the focal plane of the Flight Module No. 2 (FM2) of the JET-X telescope \citep{Citterio1994, Spiga2013}. We will show the results of the measurement campaign performed at the PANTER X-ray test facility carried out between the 27$^{th}$ of November and the 1$^{st}$ of December 2012.  
The JET-X telescope (see Tab.\ref{tab:jetx} for the characteristics) was originally built for the former SPECTRUM-X GAMMA mission.
\begin{deluxetable*}{c|c}
\tabletypesize{\scriptsize}
\tablecaption{JET-X telescope characteristics \citep{Spiga2013}.\label{tab:jetx}}
\tablewidth{0pt}
\startdata
\hline \hline
Configuration & Wolter-I\\
Focal length         & 3500 mm \\
Diameter at entrance pupil (outer shell)    &   300 mm \\
Diameter at entrance pupil (inner shell)     &   191.1 mm \\
On-axis incid. angle at the intersection plane (outer shell) & 0.60$^\circ$ \\
On-axis incid. angle at the intersection plane (inner shell) & 0.39$^\circ$ \\
Mirror length (parabolic + hyperbolic)  &   2$\times$300 mm \\
Reflecting surface material   & Gold \\
no. of shells           & 12 \\
Eff. area at 1.5 keV          & 147 cm$^2$ \\
Eff. area at 8 keV          & 53 cm$^2$ \\
FOV -- GPD+Telescope          & 14.7~arcmin~$\times$~14.7~arcmin \\ \hline
\enddata
\end{deluxetable*}

Finally, we show the simulated response for two kinds of extended sources, namely PWNe and Shell-like SNRs.
The discussion is addressed with particular emphasis with respect to the detector configuration proposed on board the small pathfinder mission XIPE (X-ray Imaging Polarimetry Explorer) \citep{Soffitta2013b} which was presented, but not selected, to the ESA call of 2012 for a small mission to be launched in 2017. Two GPDs, effective in the 2--10 keV energy band, were meant to be coupled with two JET-X optics modules to perform polarimetry of astrophysical sources.

In Sect.~\ref{sec:GPD} the GPD polarimeter and the main properties of the JET-X telescope are introduced.
In Sect.~\ref{sec:GPDFocalPlane} the arrangement of the experimental set-up is explained.
In Sect.~\ref{sec:onaxis} the on-axis angular resolution is studied, while the off-axis angular resolution is treated in Sect.~\ref{sec:offaxis}.
In Sect.~\ref{sec:opticspolarization} we discuss briefly the effects on the polarization of grazing incidence reflection of X-ray in the optics. 
In Sect.~\ref{sec:astroobservations} the implications in terms of observational targets are discussed.

\section{The GPD at the focal plane of an X-ray telescope}\label{sec:GPD}

\subsection{The GPD configuration and operation}

The GPD is a gas detector developed by the Italian research institutes INFN-Pisa and INAF/IAPS. It is designed to perform polarimetry in the X-ray energy band by exploiting the dependence of the photoelectric effect on the polarization of the radiation.
When an X-ray photon, entering through a thin Be window, is absorbed in the detector gas cell, a photoelectron is ejected and ionizes the gas atoms until it stops and releases its larger fraction of energy in the Bragg peak. The electrons of the ionization track are drifted, multiplied by a Gas Electron Multiplier (GEM) \citep{Sauli1997, Tamagawa2009}, and finally collected on a fine subdivided pixel plane (50$\mu$m of pitch). 
The analysis algorithm calculates the barycentre and the main axes of the projected charge distribution and finds the region of the track in which the projected absorption point (Impact Point--IP) is located  (oppositely to the site of the the Bragg peak) by means of a skewness analysis. The IP and the direction of ejection of the photoelectron are finally calculated as the barycentre and the major axis of the initial portion of the track, weighting properly the charge content of pixels \citep{Pacciani2003, Bellazzini2003, Soffitta2013a} due to the probable presence also of the Auger electron.

The photoelectric differential cross section for K-shell depends on the angular coordinates as follows:
\begin{equation}
\frac{d\sigma}{d\Omega}\propto \frac{\sin^2 \theta \cos^2 \phi}{(1+\beta \cos \theta)^4}\label{eq:phCrossSection}
\end{equation}
where $\beta$ is the photoelectron speed in terms of light speed units, $\phi$ is the azimuthal component of the photoelectron ejection direction and $\theta$ is the polar component. Therefore, when a polarized beam of radiation is observed, a $\cos^2 \phi$ modulation in the azimuthal distribution arises, since the photoelectrons are ejected with higher probability parallel to the X-ray photon polarization vectors. 
The energy band of the detector depends on the gas mixture composition, pressure and absorption gap thickness. It can be tuned in a range between 2 and some tens of keV, with mixtures typically of DME\footnote{DME is dimethyl ether, $\mathrm{C}_2\mathrm{H}_6\mathrm{O}$} and Helium, Neon or Argon.
The GPD collects the charge produced along the depth of the absorption gap, because the charge signal is readout from the pixel plane placed at the opposite of the Be entrance window. Therefore, photon tracks originated at different depths will suffer a different diffusion and a different recombination of the drifted ionization charge with the atoms of the gas.

The configuration operating in the 2--10 keV energy band, filled with a 20$\%$He--80$\%$DME gas mixture at 1 bar of pressure in a 1~cm thick absorption gap, equipped with a 50$\mu$m thick Be window, was the one used for the characterization at the PANTER X-ray test facility (and proposed on board the XIPE mission). This detector configuration matches very well with the typical energy range of a classical grazing incidence X-ray telescope (as JET-X).

\subsection{Imaging properties}

The angular resolution of an imaging system is limited by the blur introduced in the image of a point-like source. 
This property is summarized in the Point Spread Function (PSF), that in our case is given by the density distribution of the photon IPs on the detector image. We assumed that the PSF dependence was purely radial as modelled by \cite{Moretti2004} for the on-ground calibration of the Swift XRT telescope (mounting the FM3 mirror of JET-X). Therefore, the PSF profile is described as a Gaussian plus a King function: 
\begin{equation}
\mathrm{PSF}(r)=W e^{-\frac{r^2}{2\sigma^2}}+N\biggr(1+\biggl(\frac{r}{r_c}\biggr)^2\biggl)^{-\eta}\label{eq:PSFfunc}
\end{equation} 
We decoupled the two functions, whereas those of the original model were linked by the normalization coefficient $N$ of the King profile that was imposed equals to 1-W.
The PSF expressed as in Eq.~\ref{eq:PSFfunc} is analytically integrable in $r dr$ and its integral profile is the Encircled Energy Fraction (EEF):
\begin{widetext}
\begin{equation}
\mathrm{EEF}(r)=\int_0^r \mathrm{PSF}(r) \, 2\pi r \, \mbox{d}r=\frac{\pi r_c^2 N}{1-\eta}\biggr( \biggl(1+\biggl(\frac{r}{r_c}\biggl)^2 \biggl)^{1-\eta}-1\biggr) + 2\pi W \sigma^2 \biggl(1-e^{-\frac{r^2}{2\sigma^2}}\biggr)\label{eq:EEFfunc}
\end{equation}
\end{widetext}
so that the total flux of the source is analytically characterized:
\begin{equation}
\mathrm{EEF}(\infty)=2\pi W \sigma^2 + \pi \frac{r_c^2 N}{\eta-1} \label{eq:EEFinfty}
\end{equation}

Typically, the angular resolution is measured in terms of Half Energy Width (HEW), that analytically is defined as $\mathrm{EEF} \bigg( \frac{\mathrm{HEW}}{2} \bigg) =0.5$ for monochromatic radiation. In our case it is easy to derive the HEW as the diameter, centred around the centroid of the PSF, containing half of the IPs of the image at a given photon energy. 
It allows us to summarize, with only one parameter, the imaging performance of an optical system in terms of angular resolution. However, the accurate analysis of the PSF is needed to fully characterize the image quality, because the HEW does not take into account the PSF profile.  
The HEW was calculated by counting one by one the IPs for each image of the point-like source. The HEW is directly calculated in terms of \textit{spatial resolution} on the detector plane and then the corresponding \textit{angular resolution}  is derived taking into account the corrected\footnote{The JET-X nominal focal length is 3.5 m, but, owing to the the finite distance of the X-ray source, the true focal length is
increased to 3.6 m. The focal length, corrected for the finite distance effect is derived by the lens law $1/g+1/f=1/f_0$ where $g$ is the distance from the optics module and the source, $f$ is the corrected focal length and $f_0$ is the nominal focal length. \citep{vanSpeybroeck1972}} focal length $f=3.6$~m, following the formula:
 \begin{equation}
\mathrm{HEW[ang. \ units]}= 2 \cdot \arctan\biggl(\frac{\frac{1}{2} \cdot \mathrm{HEW[spat. \ units]}}{f}\biggr)\label{eq:calcHEWarcsec}
\end{equation}

The $1\mathrm{-}\sigma_\mathrm{HEW}$ errors are also calculated, according to the binomial statistics.
Since the HEW contains half of the counts of a PSF image, the probability for an IP to stay within the HEW is $p=0.5$, therefore the binomial fluctuation associated to the number of events within the HEW is given by:
\begin{equation}
\sigma_{ct}=\sqrt{var_{ct}} =\sqrt{n\cdot p(1-p)}=\frac{1}{2}\sqrt{n}\label{eq:sigmabinomial}
\end{equation}
where $n$ is the total number of counts.
The $\sigma_\mathrm{HEW}$ is derived by calculating the HEW for the fluctuation of count corresponding to $\frac{1}{2}n\pm \sigma_{ct}$. 
Therefore, the statistical fluctuation in terms of counts corresponds to a statistical fluctuation in terms of HEW.

\begin{figure} 
 \begin{center}
\begin{tabular}{c}
\includegraphics[scale=0.45]{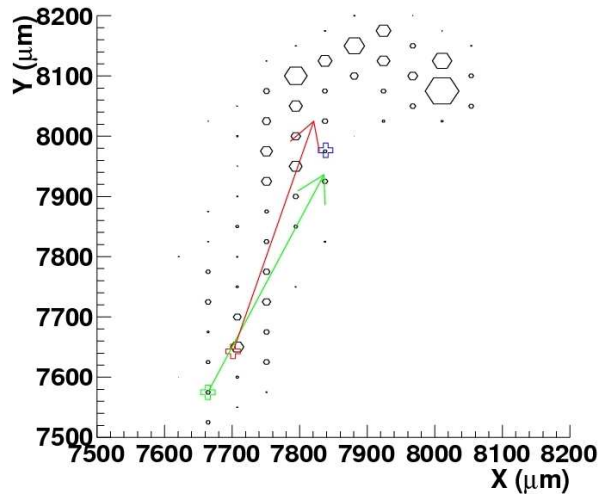}
\end{tabular}
\caption{Simulated phtoelectron track produced by the absorption of an 8 keV photon in a gas mixture given by He20$\%$--DME80$\%$ at 1 bar of pressure. The hexagons represent the pixels and the size is proportional to the amount of charge collected. The empty crosses represent the barycentre of the charge distribution (blue), the true absorption point (green) and the reconstructed IP (red). The true photoelectron ejection direction is represented by a green arrow, while the reconstructed one is coloured in red.}\label{fig:pixmap}
 \end{center}
 \end{figure}

The angular resolution of the system composed by the GPD and an X-ray telescope has three different contributions.
The first contribution is due to the optics PSF and results in a spread of photons on the focal plane, because rays in gas are deviated from the ideal focusing. The second contribution derives from the inclined penetration and absorption of photons through the thickness of the GPD absorption gap \citep{Lazzarotto2010}. In past years this effect was called with the \textit{ambiguous} terms ``parallax"  \citep{Gabriel1977, Lewis1994}.
This effect causes a small PSF degradation (few arcseconds) with respect to the intrinsic telescope angular resolution due to the fact that the penetration angle in gas, with respect to the mirror module focal axis, is small. It amounts to 4 times the angle of incidence of radiation on the mirror shells and varies from 0.60$^\circ$ for the most external shell down to 0.39$^\circ$ for the innermost one.
The third contribution to the angular resolution is given by the intrinsic spatial resolution of the detector that depends on the shape of the photoelectron tracks which in turn depends on: 
\begin{enumerate}
\item scattering behaviour of ejected photoelectrons
\item diffusion properties of the ionization charge (gas mixture diffusion coefficient, pressure, drift length)
\item detector pixel size (50~$\mu$m of pitch)
\end{enumerate}
These effects impact on the accuracy of the IP measurement performed by the track reconstruction algorithm. The contribution of the intrinsic spatial resolution of the GPD is shown in Fig.~\ref{fig:pixmap}, where we show a simulated phtoelectron track produced by a photon of 8 keV of energy. In this figure, the hexagons represent the pixels and their size is proportional to the amount of charge collected. The empty crosses represent the barycentre of the charge distribution (blue), the projection onto the pixel plane of the true absorption point (green) and the reconstructed IP (red). Also the projection of the true photoelectron ejection direction (green arrow) and the reconstructed one (red arrow) are shown. The capability to reconstruct the impact point (red cross) as close as possible to the true absorption point (green cross) depends on the intrinsic detector spatial resolution.

\section{Configuration of the set-up}\label{sec:GPDFocalPlane}

 \begin{figure} 
 \begin{center}
\begin{tabular}{c}
\includegraphics[scale=0.15]{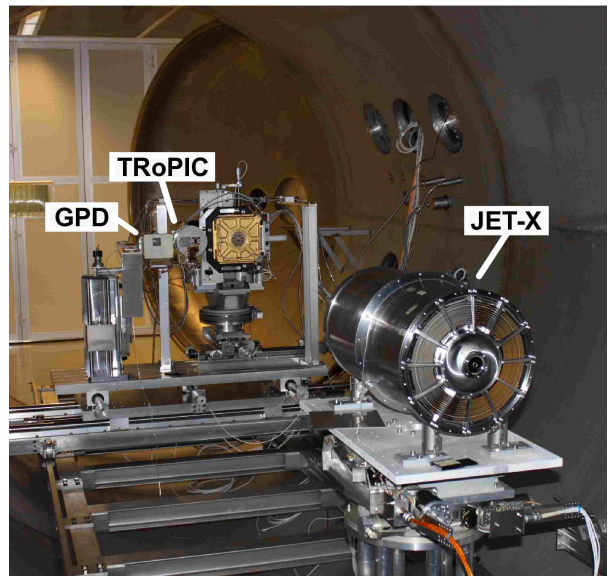}
\end{tabular}
\caption{Measurement set-up in the vacuum chamber at PANTER.}\label{fig:SetUp}
 \end{center}
 \end{figure}
\begin{figure} 
 \begin{center}
\begin{tabular}{c}
\includegraphics[scale=0.55]{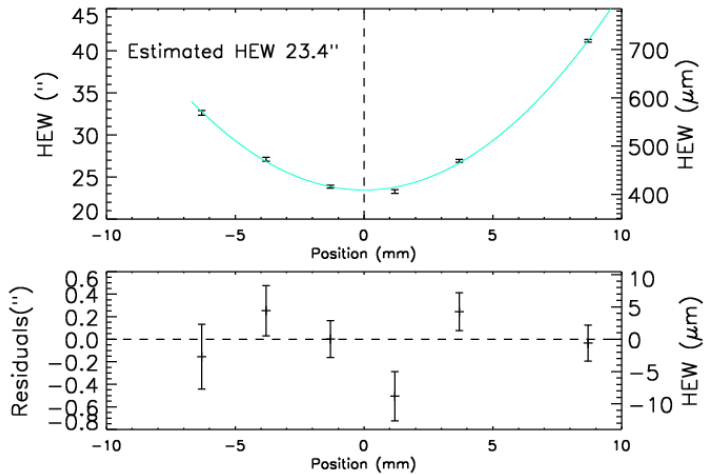}
\end{tabular}
\caption{Each point represents the HEW calculated for different distances between the GPD and the telescope optics. By fitting a parabola on the points near the minimum, the position corresponding to the better on-axis angular resolution at 4.51 keV is obtained. The mirror module was shifted by steps of 2.25$\mu$m of accuracy. Moving from left to right on the abscissa the detector/mirror separation increases.}\label{fig:BestFocusSearch}
 \end{center}
 \end{figure}

The PANTER X-ray test facility is described in \cite{Freyberg2005} and \cite{Burwitz2013}. We installed the GPD in the test room (see Fig.~\ref{fig:SetUp}) and firstly searched for the best focus position after centring the X-ray spot on the detector active area. We shifted the X-ray mirror along its optical axis to find the position where we obtained the best angular resolution at the energy of 4.51 keV. 
This energy is about in the middle of the sensitive band and it is well explored in our laboratory.
For all measurements described in this work the background is negligible and no subtraction to the source signal was needed.
Fig.~\ref{fig:BestFocusSearch} shows the HEW measured for different distances between the GPD and the telescope optics. 
By fitting a parabola on these points, the position corresponding to the minimum HEW on axis is located. 
The characterization measurements for the detector and the optics were performed at the energies of 2.98, 4.51 and 8.05 keV, corresponding respectively to fluorescence lines of Ag (L), Ti (K) and Cu (K) targets, excited by accelerated electrons in the facility radiation source, which was located 128~m far from the mirror module in the test room \citep{Spiga2013}. The radiation beam was filtered to remove the bremsstrahlung continuum out of the spectral region of the fluorescence lines. Low energy wings, albeit small, due to the incomplete charge collection of photoelectron tracks exiting from the edges of the detector active volume were still present in the spectra. They were removed via software by selecting only events corresponding to the photo-peak in each spectrum.
No other selection cuts were applied to the photoelectron tracks.

\begin{figure} 
 \begin{center}
\begin{tabular}{c}
\includegraphics[scale=0.55]{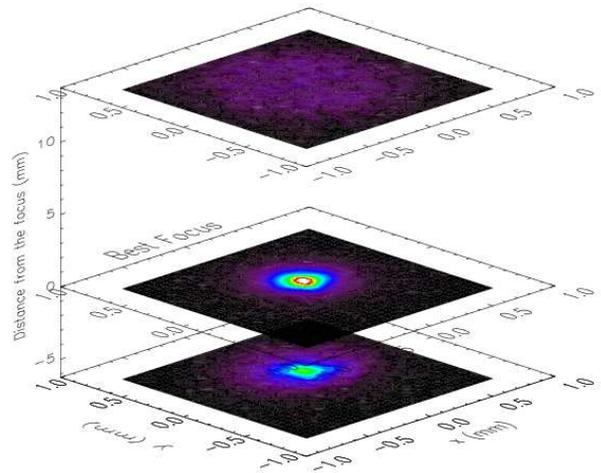}
\end{tabular}
\caption{IP maps obtained for three different distances between the GPD and the telescope optics. The plots are normalized to the number of counts for each image. At the central position the image corresponding to the better angular resolution, is shown. A narrower PSF core (white spot) with respect to the other images is present in this one.}\label{fig:ThreePositions}
 \end{center}
 \end{figure}

In Fig.~\ref{fig:ThreePositions} the IPs maps corresponding to the position where the HEW at 4.51 keV is minimized and two displaced positions are shown. The plots are normalized to the number of counts for each image. It is visible, even at first glance, that the central image has a more pronounced PSF core (white spot) with respect to the other images, whose counts are more spread.

\section{On-axis angular resolution}\label{sec:onaxis}

\begin{figure*} 
 \begin{center}
\begin{tabular}{c}
\includegraphics[scale=0.45]{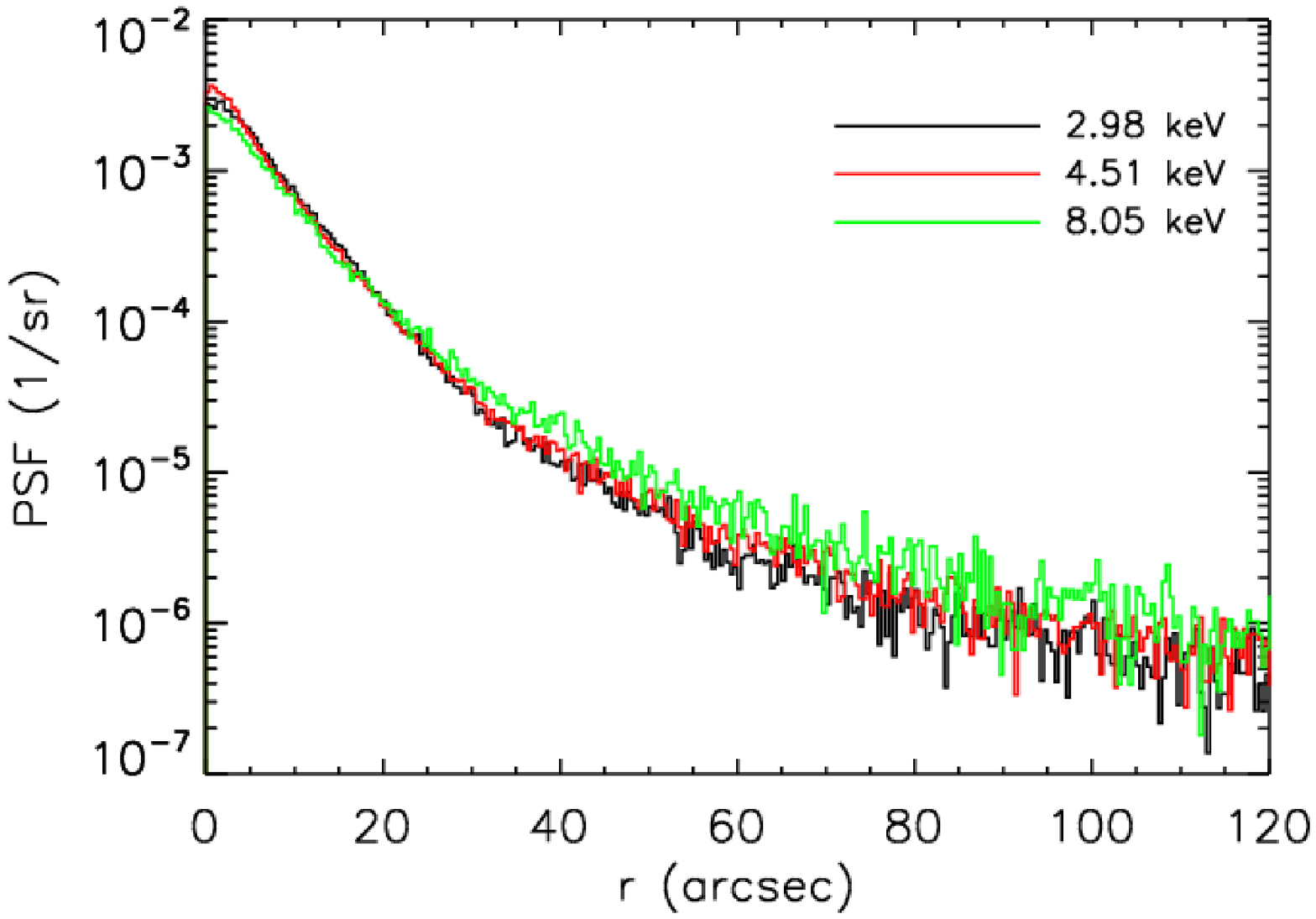}
\includegraphics[scale=0.45]{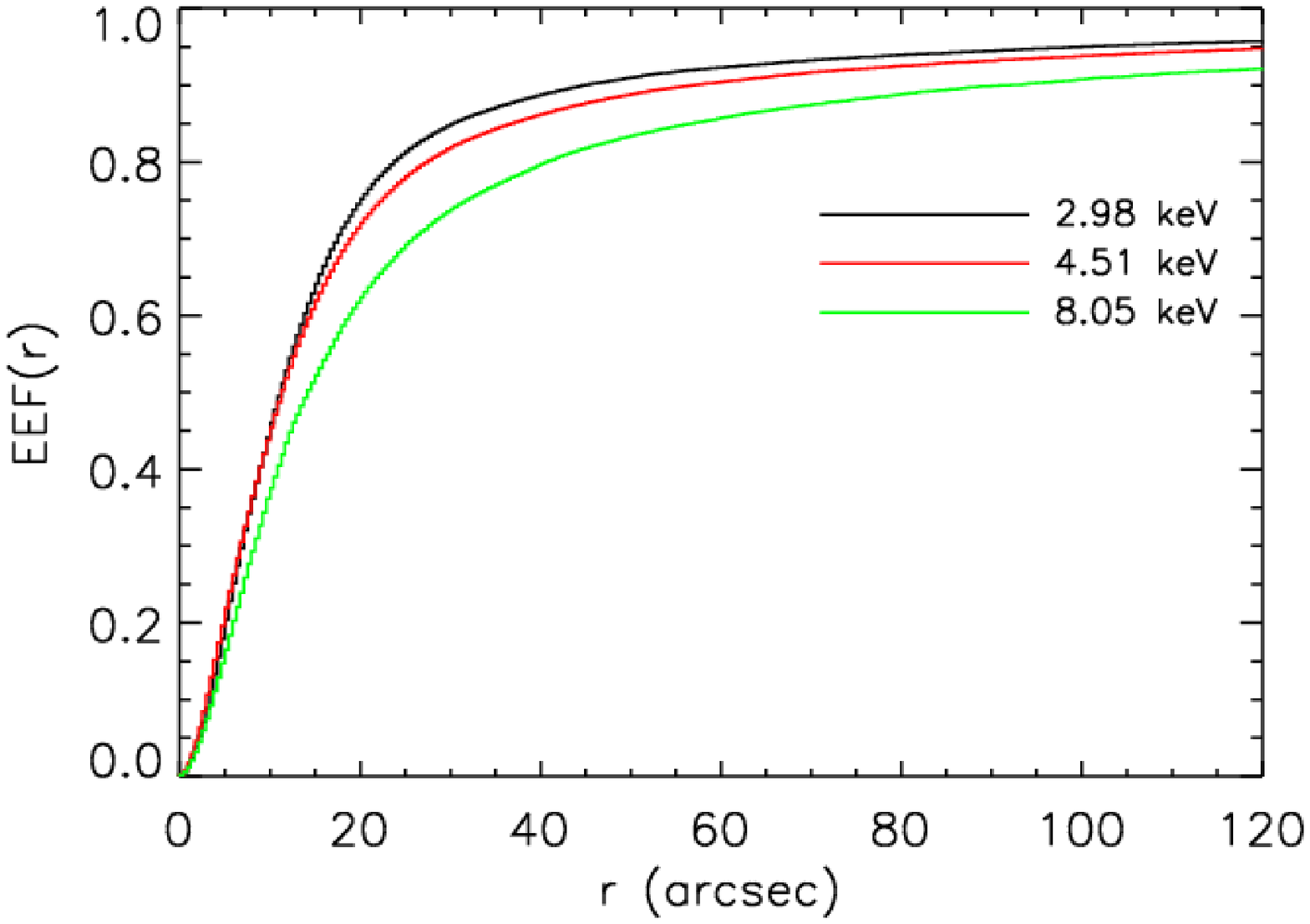}
\end{tabular}
\caption{On the left: comparison of the $PSF(r)$ profiles of the 2.98, 4.51 and 8.05 keV radiation beams. The profiles are normalized to the total number of counts.  At higher energies the wings contribution is larger. On the right: the corresponding $EEF(r)$ derived from the data. The contribution of wings keeps the $EEF$ well below the unity even for large $r$ values, in particular at higher energies. }\label{fig:PSFconfrontation}
 \end{center}
 \end{figure*}

\begin{figure}
 \begin{center}
\includegraphics[scale=0.35,angle=-90]{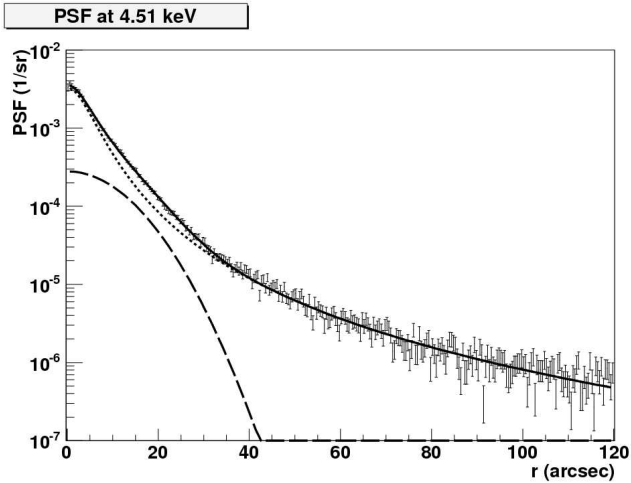}
\caption{Fit performed on the PSF at 4.51 keV with a Gaussian plus a King function. The fit is performed on the radial coordinate $\theta$ from 0 to 120~arcsec. See Tab.~\ref{tab:PSFFit} for the fit results at the energy of 2.98, 4.51 and 8.05 keV. Each fit is performed on the radial coordinate $\theta$ from 0 to 120 arcsec and the IPs density distribution is normalized to the total number of counts.
}\label{fig:PSFFit}
 \end{center}
 \end{figure}

\begin{figure*} 
 \begin{center}
\begin{tabular}{c}
\includegraphics[scale=0.9]{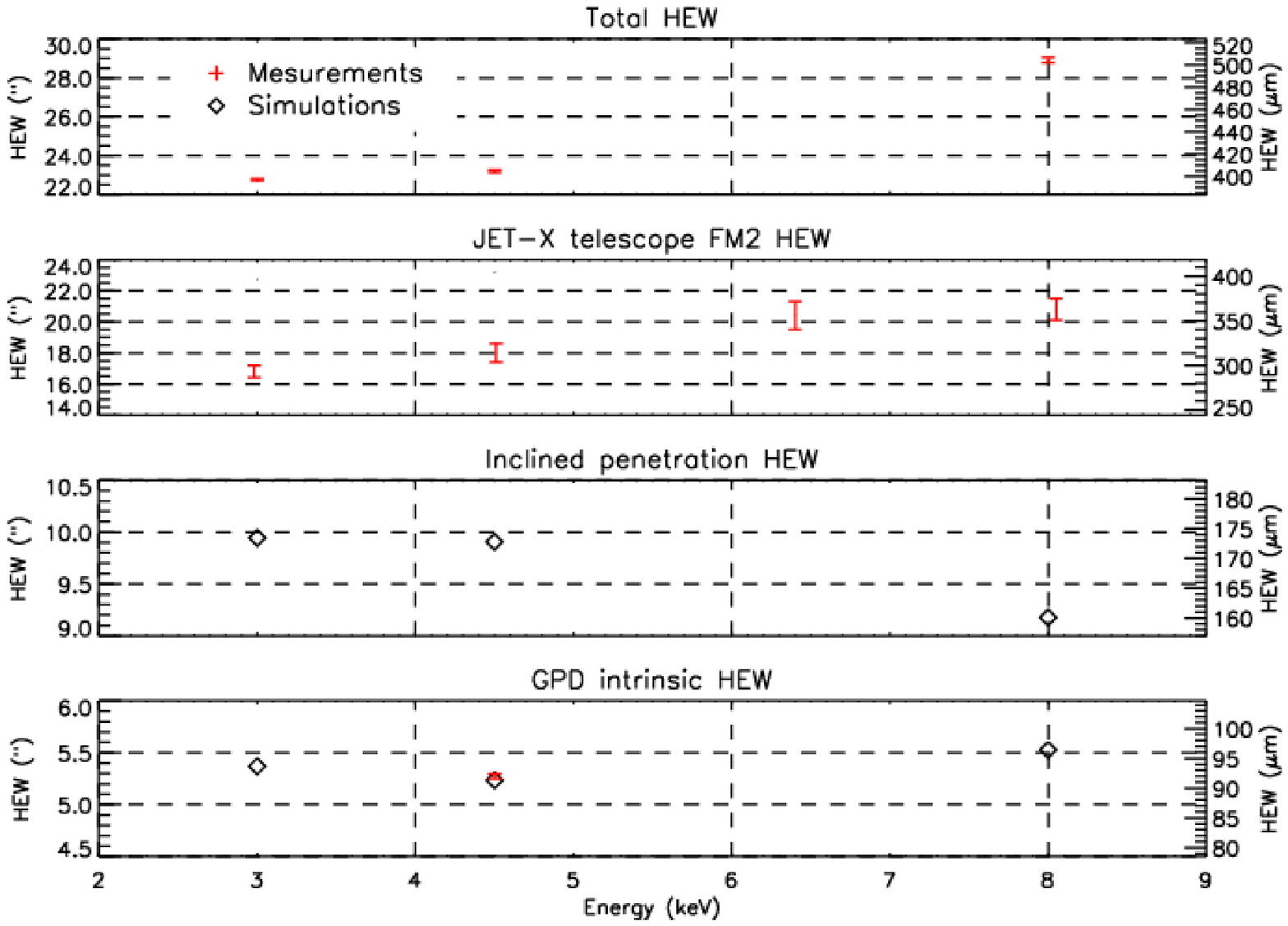}
\end{tabular}
\caption{HEW at 2.98, 4.51 and 8.05 keV for the on-axis configuration for the three different components that contribute to the overall PSF of the GPD/X-ray telescope optical system. 
In the top panel the HEW measured at the PANTER X-ray test facility of the GPD coupled at the JET-X FM2 optical module.
Plotted values are listed in Tab.~\ref{tab:HEWoffAxis}.
In the second panel (from the top) is shown the intrinsic telescope HEW \citep{Spiga2013} for a 2~cm$^2$ active area detector.
In the third panel the HEW simulated for the inclined penetration of radiation in gas by assuming the ideal optics case, therefore without the contribution of the intrinsic PSF of the optical module. Also the effect of the intrinsic spatial resolution of the detector is not taken into account. This plot shows the contribution of the simple geometrical model of the optics to the overall PSF. Radiation is assumed to come from infinity, therefore the simulated focal length is 3.5~m. 
In the bottom panel the HEW for a zero-width vertical beam of radiation is shown both from simulations and from one measurement at 4.51 keV. (red, data re-analysed from \cite{Soffitta2013a}). The measurement result is obtained by quadratically subtracting the intrinsic beam HEW.}\label{fig:HEWonAxis}
 \end{center}
 \end{figure*}

The on-axis angular resolution was measured also at 2.98 and 8.05 keV with the detector in the position where the on-axis angular resolution at 4.51 keV was minimized.
The profile of the overall PSF at 2.98, 4.51 and 8.05 keV of the optical system given by the JET-X FM2 plus the GPD is shown on the left panel of Fig.~\ref{fig:PSFconfrontation}, on the right panel the corresponding EEF profile is shown.
In Fig.~\ref{fig:PSFFit} is shown the fit with the PSF function of Eq.~\ref{eq:PSFfunc} on the measured IPs density distribution (normalized to the total number of counts) for radiation at 4.51 keV. In Tab.~\ref{tab:PSFFit} the fit parameters for the 2.98, 4.51 and 8.05 keV energies are listed. 
In Fig.~\ref{fig:HEWonAxis} are shown the measured values of the HEW (top panel). They are also listed in Tab.~\ref{tab:HEWoffAxis}. 
\begin{deluxetable*}{ccccc}
\tabletypesize{\scriptsize}
\tablecaption{Results of the fit performed on the PSF, normalized to the total number of counts, with a Gaussian plus a King function shown in Fig.~\ref{fig:PSFFit} only for 4.51 keV. Each fit is performed on the radial coordinate $\theta$ from 0 to 120 arcsec. \label{tab:PSFFit}}
\tablewidth{0pt}
\tablehead{& &\colhead{2.98 keV } & \colhead{4.51 keV} & \colhead{8.05 keV}}
\startdata
   \colhead{$\chi^2/\mathrm{ndf}$} &  &342.5/283 & 322.9/283 &  303.3/283 \\ 
  \colhead{ $\chi^2$ Norm.} &  &1.21 & 1.14 &  1.07\\  
 \colhead{ $W\pm\sigma_W$} &  (1/sr) &$(3.87 \pm 0.027)\times10^{-4}$  &  $(2.79\pm0.21)\times 10^{-4}$  & $(5.5 \pm 1.4)\times10^{-4}$ \\
 \colhead{ $\sigma\pm\sigma_\sigma$ } & (arcsec) &  $9.85 \pm 0.13$ &  $10.61\pm0.16$ &  $2.77\pm0.51$ \\
 \colhead{ $N\pm\sigma_N$} & (1/sr) &  $(2.574\pm 0.049)\times10^{-3}$ &  $(3.2890\pm0.0057)\times 10^{-3}$  & $(1.94\pm0.14)\times 10^{-3}$ \\
\colhead{$r_c\pm\sigma_{r_c}$ } &  (arcsec) & $7.57 \pm 0.18$  & $6.06\pm0.14$  &  $9.57\pm0.39$ \\ 
 \colhead{$\eta\pm\sigma_\eta$} & &  $1.629\pm 0.019$ &  $1.481\pm0.014$  &  $1.606\pm0.020$\\\hline 
\enddata
\end{deluxetable*}

As discussed in Sect.~\ref{sec:GPD}, the angular resolution of the GPD coupled to the telescope optics is affected by three blurring effects: the intrinsic PSF of the optics, the blurring induced by inclined penetration of photons in gas and the uncertainty in the determination of the photon IPs due to the intrinsic detector spatial resolution. While the first effect is an intrinsic property of the telescope optics alone, the second depends both on the optics and on the detector parameters, and the third one depends on the detector parameters and on the analysis procedure. 
We are interested in decoupling each contribution with respect to the total angular resolution. 
The intrinsic telescope angular resolution in terms of HEW is shown in Fig.~\ref{fig:HEWonAxis} (second panel from the top).  It was recently measured at the PANTER X-ray test facility \citep{Spiga2013} at the energy of 1.49 (Al-K), 2.98, 4.51, 6.4 (Fe-K) and 8.05 keV. The measurement was performed with the TRoPIC CCD \citep{Predehl2007} placed at the focal plane of the JET-X FM2. The area of the TRoPIC detector (1.96~cm~$\times$~1.96~cm) is nearly 4~cm$^2$, therefore larger than the one of the GPD (2.25~cm$^2$).
Nevertheless, the FM2 PSF is so compact that its wings are almost entirely included in a 2~cm$^2$ region, at least in the range of measured X-ray energies, so the exact size of the GPD active region is not crucial.
The GPD condition is therefore well reproduced adopting the HEW values 
computed over a 2~cm$^2$ area as done by \cite{Spiga2013}.

\begin{figure*} 
 \begin{center}
\begin{tabular}{c}
\includegraphics[scale=0.3]{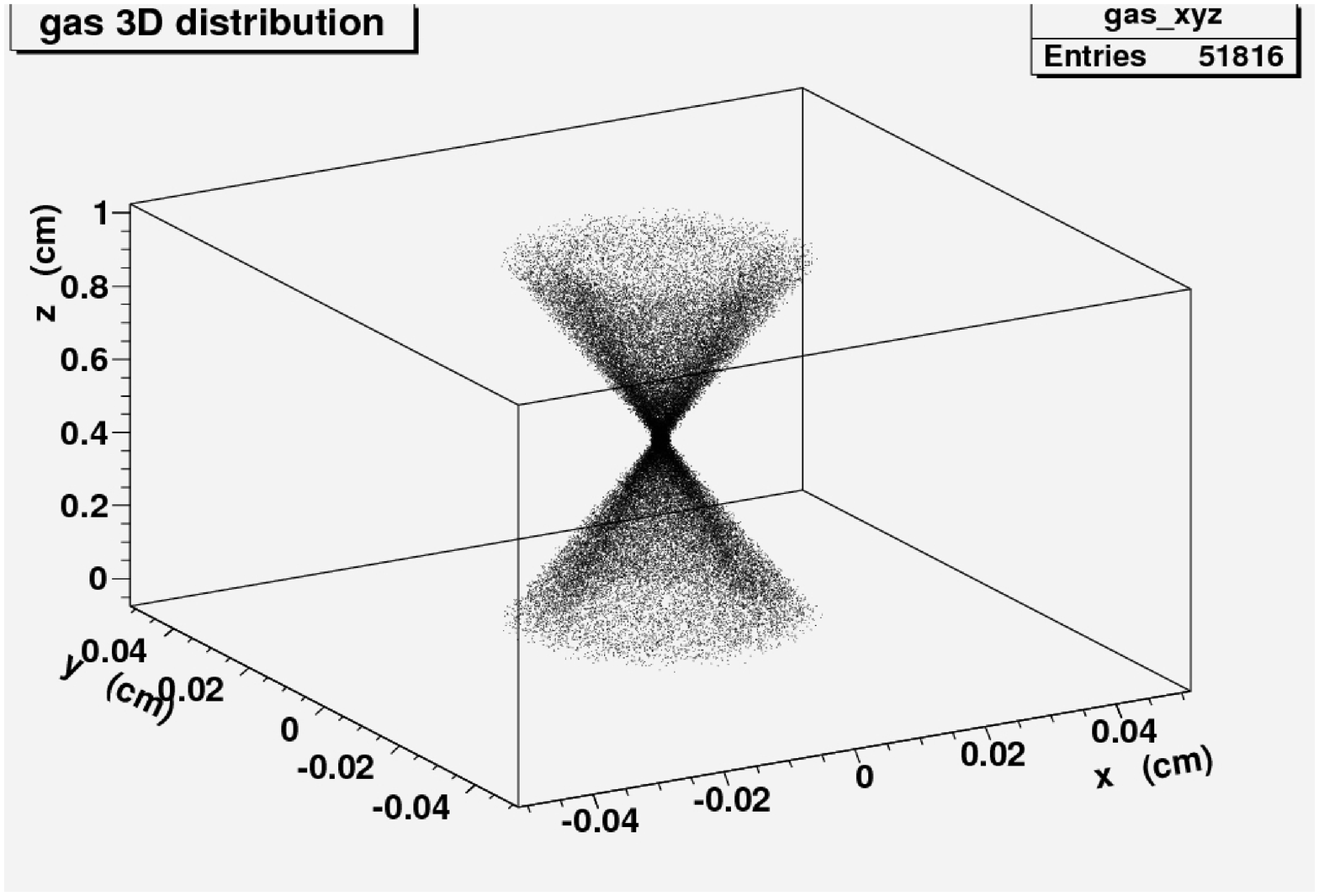}
\includegraphics[scale=0.3]{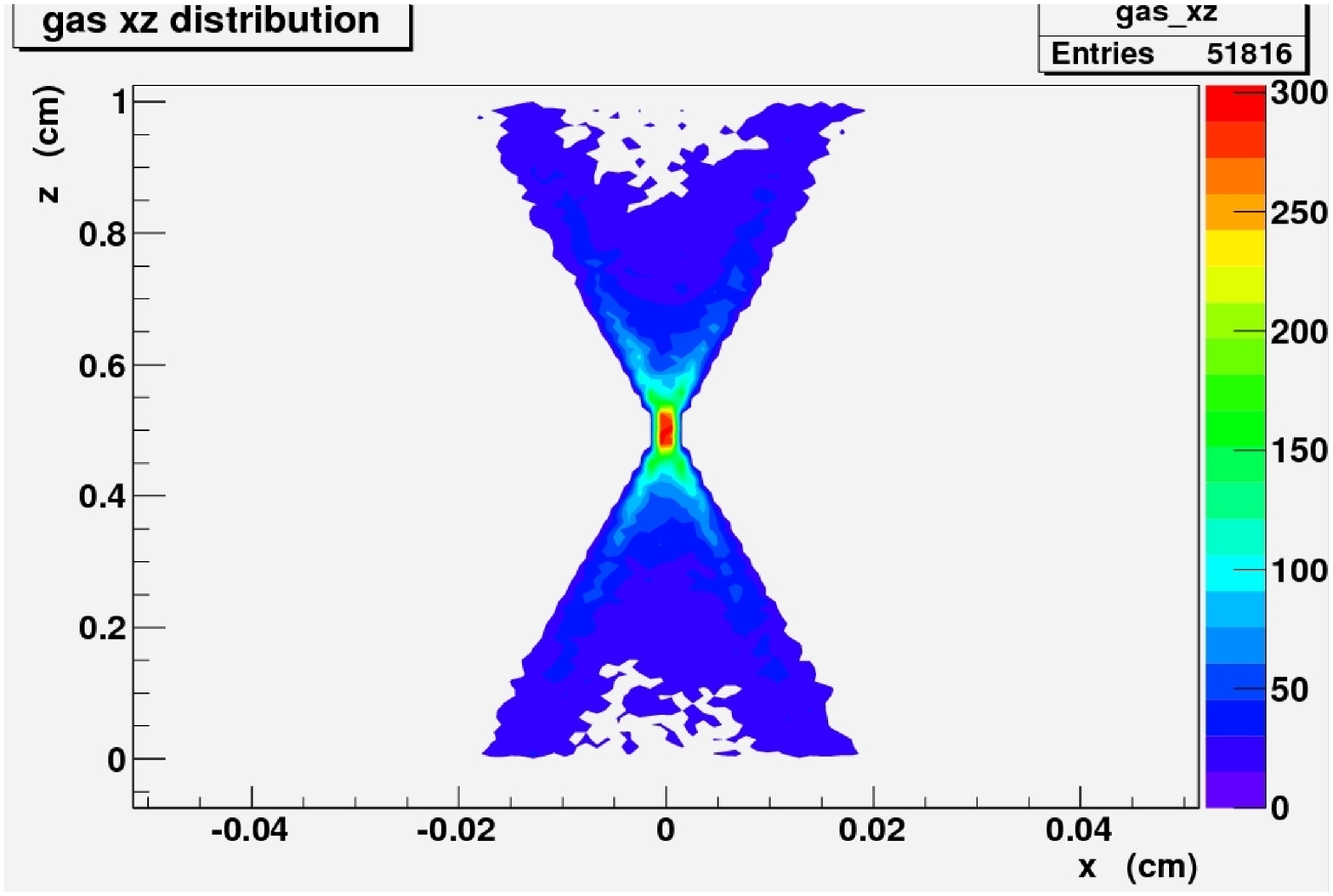}\\
\includegraphics[scale=0.3]{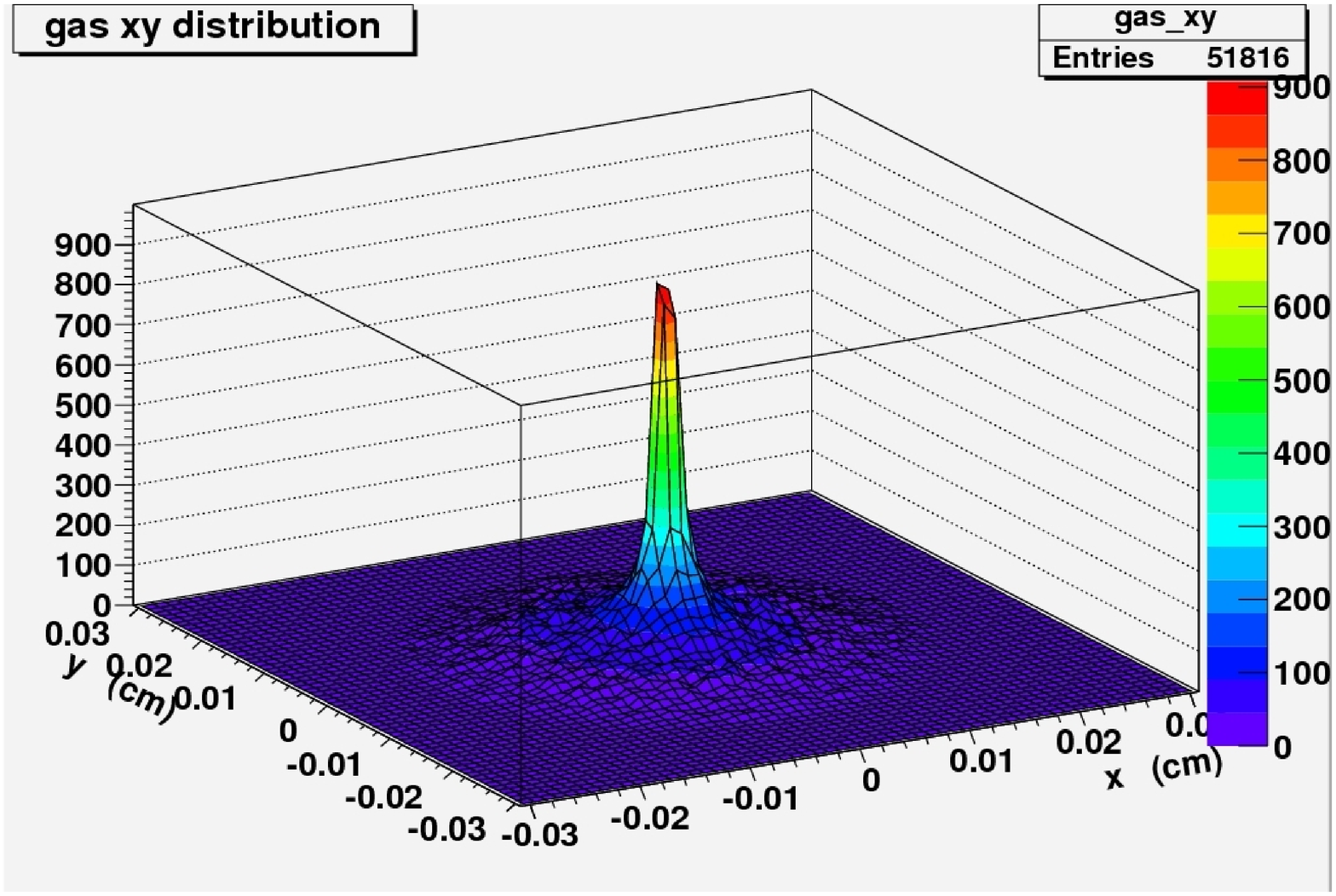}
\includegraphics[scale=0.3]{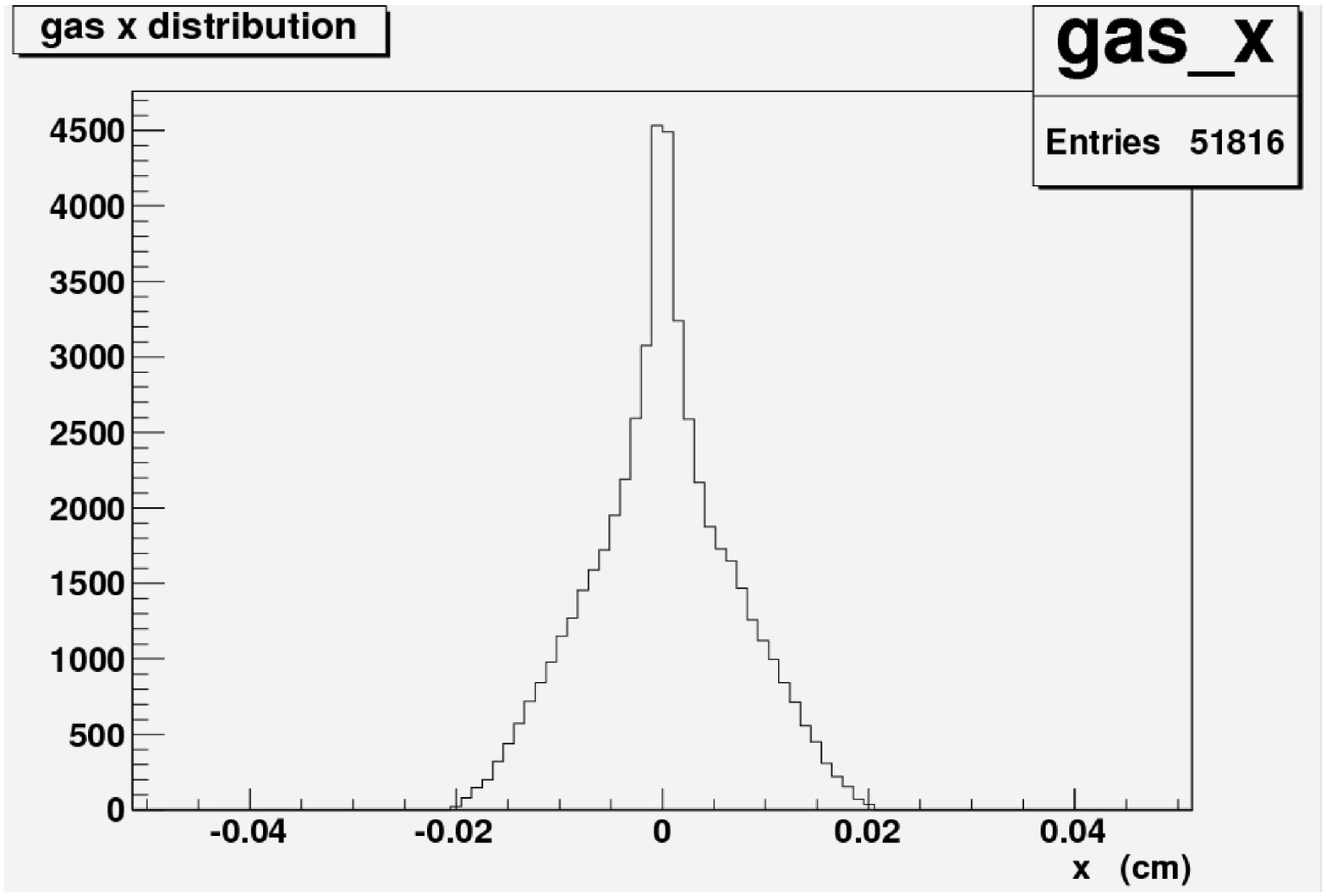}
\end{tabular}
\caption{Simulation for radiation at 4.5 keV of the absorption points distribution due to inclined penetration only for the detector with 1~cm of absorption gap filled with 20$\%$He--80$\%$DME at 1 bar of pressure. The top left: 3D distribution of absorption points. The coordinate $z=0$ is the detector top face (Be window). Top right panel: The $zx$ marginal distribution. Bottom left panel: distribution on the $xy$ plane. Bottom right panel: marginal distribution along $x$ axis.}\label{fig:InclinedSim}
 \end{center}
 \end{figure*}

The contribution of the inclined penetration of photons in gas is evaluated by means of simulations \citep{Lazzarotto2010}, assuming that the photons propagate along ideal reflection paths, therefore neglecting the telescope PSF and the intrinsic GPD spatial resolution.
In this simulation the radiation is assumed to come from infinity, therefore the nominal focal length of 3.5~m is assumed.
In Fig.~\ref{fig:InclinedSim} we show the simulation of the distribution of the absorption points. 
The 3D distribution is in the top left panel, whereas the marginal $zx$ and $xy$ planes are reported in the top right and bottom left panels, respectively.
Eventually, the marginal distribution along $x$ axis is reported in the bottom right panel. Due to the focusing and the absorption along the depth of the gas cell, the distribution of the absorption points on the readout plane is characterized by a narrow core and extended wings. The distribution of absorption point on the $xy$ plane, that corresponds to the readout plane, is not Gaussian.
The results of the simulations in terms of HEW are plotted in the third panel from the top of Fig.~\ref{fig:HEWonAxis}. It is important to notice that in the case of the GPD, having a 1~cm thick gas cell, this effect contributes for about 10~arcsec (in coupling with JET-X optics) to the total angular resolution.
This term of the angular resolution is better for higher energies, since high energy radiation is efficiently reflected by more internal mirror shells, so that it penetrates with smaller inclination angles with respect to the optical axis.

Finally, the blurring due to the intrinsic spatial resolution of the detector is reported in terms of HEW in the bottom panel of Fig.~\ref{fig:HEWonAxis}. This component was studied by \cite{Soffitta2013a} that represented the impact point distribution with a bivariate Gaussian profile. Here we re-analysed their results to determine the angular resolution in terms of HEW (not sigma as done by \cite{Soffitta2013a}).
In the bottom panel of Fig.~\ref{fig:HEWonAxis} the simulated HEW (black diamonds) is evaluated considering photons impinging normally on the detector with a beam of intrinsic zero width. At the energy of 4.51~keV the simulation is compared with the result of an experimental measurement with a real vertical beam with a finite (albeit small) size. 
To compare properly the simulation with the measurement the intrinsic width of the radiation beam was measured in terms of  $\sigma_x=14.7 \ \mu\mathrm{m}$ and  $\sigma_y=8.7 \ \mu\mathrm{m}$ by scanning the beam with a SiPIN detector coupled with a slit (see \cite{Soffitta2013a} for details). The average sigma was calculated by approximating the beam profile with a circularly symmetric Gaussian bivariate distribution:
$f(r)= \frac{1}{2\pi \sigma^2} e^{(-\frac{r^2}{2 \sigma^2})} 2 \pi r$
with $\sigma=(14.7+8.7)/2=11.7 \mu\mathrm{m}$. The corresponding $\mathrm{HEW_{beam}}=27.6 \ \mu\mathrm{m}$  was derived according to the formula
\begin{equation}
0.5=\int_{0}^{r_0} f(r) dr =1- e^{-\frac{r_0^2}{2\sigma^2}} \label{eq:hewintsub}
\end{equation}
where 0.5 is the fraction of events inside a circle of radius $r_0=\frac{1}{2}\mathrm{HEW_{beam}}$

The corresponding $\mathrm{HEW_{beam}}$ of the beam was quadratically subtracted from the $\mathrm{HEW_{gross}}=96.1 \ \mu\mathrm{m}$ calculated from the measurement performed with the GPD to derive the intrinsic HEW of the detector that is $92.1 \ \mu\mathrm{m}$.

Some concern could arise, with respect to the measurement of polarization, from the small difference of the sigma of the Gaussian bivariate distribution that describes the intrinsic GPD response found by \cite{Soffitta2013a}. This is possibly due to the different X and Y sampling in a 50~$\mu$m pitch hexagonal pattern of pixels. One could speculate that this affects the determination of the ejection direction of photoelectrons, because it is derived by means of a statistical analysis on the spatial distribution of charge. However, this difference is small, even smaller than the pixel pitch.
Therefore, the effect on the polarimetric capability is negligible, if present, because the spurious polarization of the GPD alone (measured with unpolarized radiation) is ($0.18 \pm 0.14$)~$\%$ \citep{Bellazzini2010}. This means that no evident spurious effects are detected depending on the honeycomb pattern of hexagonal pixels.

The vertical beam geometry approximates reasonably the case of penetration of radiation with small inclination with respect to the optics module axis.
In fact, since photoelectrons are ejected with higher probability on the plane normal to the photon incident direction (see Eq.~\ref{eq:phCrossSection}), if the radiation beam is inclined by a small angle (as in the case of the JET-X telescope) this plane is quite parallel to the pixel plane. 
In case of narrower penetration angles, as it would be for longer focal lengths and same optics diameter, the energy band of the telescope would be extended to higher energies and the blurring due to inclined penetration of photons in gas would be reduced. On the other hand, with respect to the intrinsic spatial resolution of the GPD the beam geometry would tend to approximate more closely the vertical beam condition for which the intrinsic GPD spatial resolution was simulated and measured (see bottom panel of Fig.~\ref{fig:HEWonAxis}).

Looking at all panels of Fig.~\ref{fig:HEWonAxis}, it turns out that the overall HEW increases with energy (top panel) and the larger contribution to this increase comes from the intrinsic PSF of the mirror module (second panel from the top).
Since some blurring component does not follow a pure Gaussian statistics, the HEW of all components can not be properly summed in quadrature. As first approximation the Gaussian statistics can describe only the core of the response of the detector to a vertical narrow beam \citep{Soffitta2013a}.

Results reported confirm the GPD to be a high angular resolution focal plane instrument, since the performance of the telescope are not compromised by coupling it with the 1~cm thick gas detector.

\section{Off-axis angular resolution}\label{sec:offaxis}

\begin{figure}
\begin{center}
\includegraphics[scale=0.38,angle=-90]{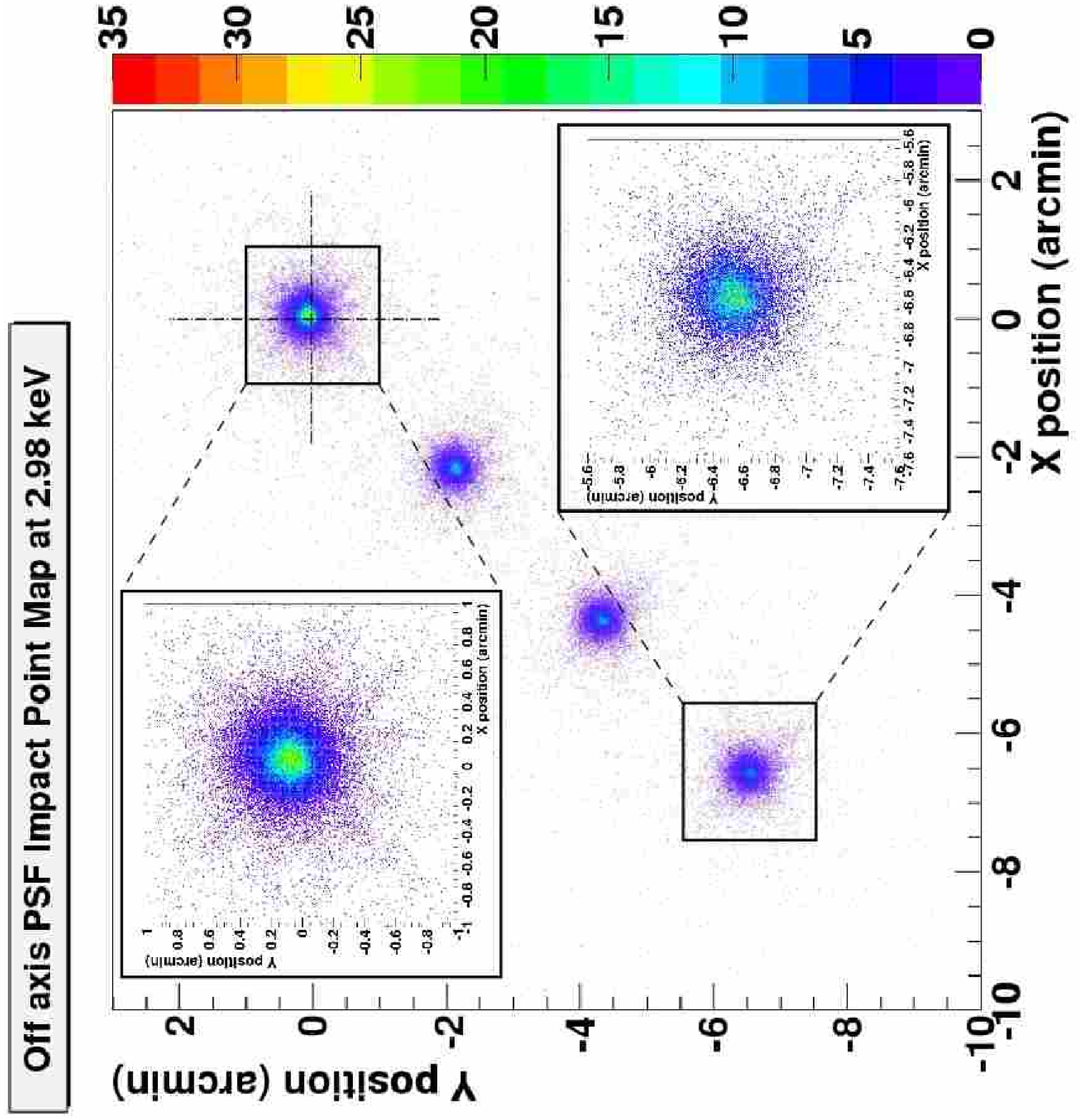}\\
\includegraphics[scale=0.38,angle=-90]{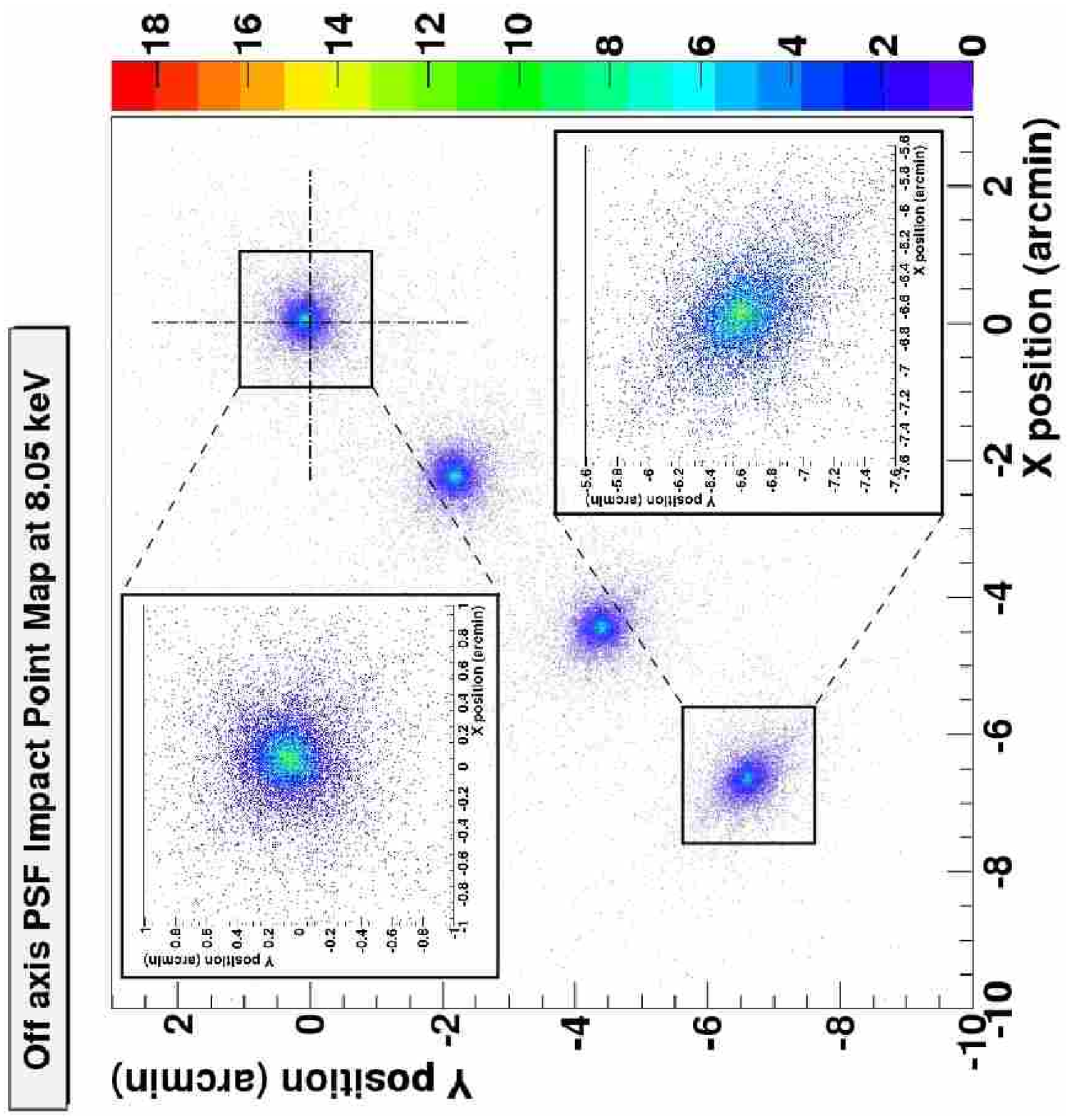}
\end{center}
\caption{Impact Point maps of the off axis measurements at 2.98 (top panel) and 8.05 keV (bottom panel). The corresponding HEW values (also for 4.51 keV) are listed in Tab.~\ref{tab:HEWoffAxis} and shown in Fig.~\ref{fig:HEWoffAxis}. 
In the two panels the zooms of the on axis measurement (the FOV center is identified by a dashed cross) and of the larger off axis 
measurement are shown.
\label{fig:HEWoffAxisImpactPointMap}}
\end{figure}

\begin{figure}
\epsscale{1.3}
\plotone{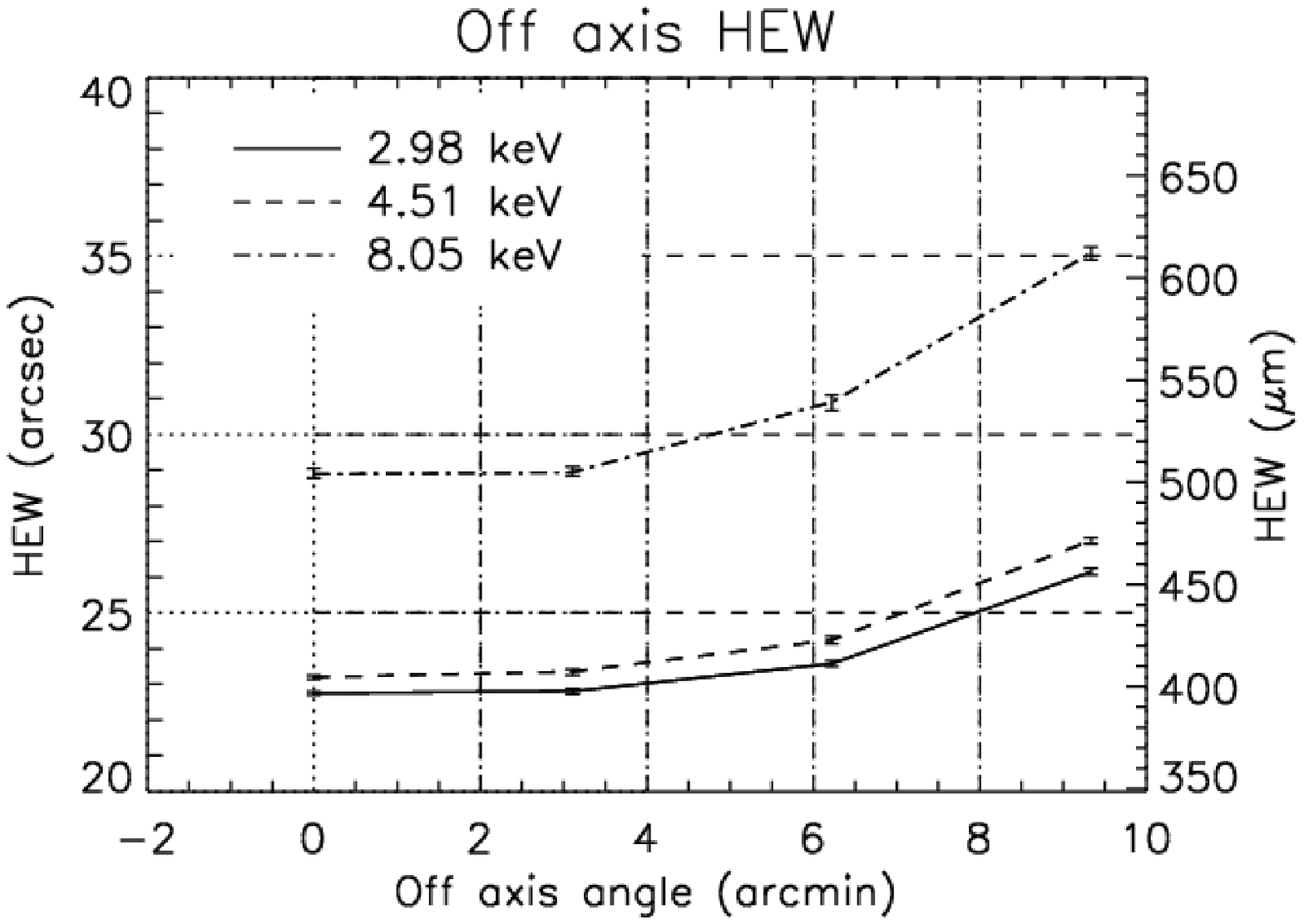}
\caption{Off axis HEW at 2.98, 4.51 and 8.05 keV. Plotted values are listed in Tab.~\ref{tab:HEWoffAxis}. The FOV of JET-X coupled with the GPD (see Tab.\ref{tab:jetx}) corresponds to an angle of 10.4~arcmin from the image center to the corner along the diagonal. \label{fig:HEWoffAxis}}
\end{figure}

The angular resolution for radiation beams impinging off-axis on the telescope was measured at 2.98, 4.51 and 8.05 keV. This characterization is crucial to verify the telescope/detector response for extended sources, whose radiation come also from directions different from the optical axis. 
The measurements were performed by tilting the mirror module in azimuth and polar angle, to reproduce the effect of observing a point source in the different regions of the FOV . The radiation beam impinged always in the same central position on the detector plane, but the PSF obtained corresponded to offset images along the diagonal of the FOV in the third quadrant at 3.11~arcmin, 6.22~arcmin and 9.33~arcmin from the central position (see Fig.~\ref{fig:HEWoffAxisImpactPointMap}). The effect of the inclined penetration due to the off-axis angle is negligible with respect to the blurring induced by the inclined penetration due to the beam focusing.  
In Tab.~\ref{tab:HEWoffAxis} the HEW values for the off-axis measurements are listed and in Fig.~\ref{fig:HEWoffAxis} they are shown in the same plot.
\begin{deluxetable*}{crrc}
\tabletypesize{\scriptsize}
\tablecaption{Angular resolution of the GPD coupled with the JET-X FM2 optics measured at 2.98, 4.51 and 8.05 keV both for on-axis and off-axis radiation beams. Listed values are plotted in Fig.~\ref{fig:HEWonAxis} (on-axis radiation beams as a function of energy) and Fig.~\ref{fig:HEWoffAxis} (angular resolution at different energies as a function of the off-axis angle). 
 \label{tab:HEWoffAxis}}
\tablewidth{0pt}
\tablehead{\colhead{Energy (keV)} & \colhead{Off Axis Angle (arcmin) } & \colhead{HEW (arcsec) } & \colhead{$\pm \sigma_{\mathrm{HEW}}$ (arcsec) }}
\startdata
\hline
2.98 keV& &  &  \\ \hline
&0.00 & 22.7 & $\pm 0.1$ \\
&3.11	& 22.8	& $\pm 0.1$\\
&6.22 &	23.6	& $\pm 0.1$\\
&9.33 &	26.1	& $\pm 0.1$\\ \hline \hline
4.51 kev& &  &  \\ \hline
&0.0 0 & 23.2 &	$\pm 0.1$\\
&3.11	&23.4&	$\pm 0.1$\\
&6.22	&24.2&	$\pm 0.1$\\
&9.33	&27.0&	$\pm 0.1$\\ \hline \hline
8.05 kev& &  &  \\ \hline
&0.00  & 28.9 &	$_{-0.1}	^{+0.2}$\\
&3.11 &	28.9 &	$_{-0.1} 	^{+0.2}$\\
&6.22 &	30.9 &	$\pm 0.2$\\
&9.33 &	35.1 &	$\pm 0.2$\\
\enddata
\end{deluxetable*}

The PSF deformation, due to the off axis imaging, is small even for the 8.05 keV radiation beam at the larger off axis angle sampled that is 9.33~arcmin. The off axis measurements confirm the capability of the GPD to perform imaging also for extended sources like PWNe and SNRs.

\section{Polarization and grazing incidence in X-ray telescopes}\label{sec:opticspolarization}

The assessment of the polarization introduced by the optics is an important point when dealing with detectors that measure the polarization.
We know by theory \citep{Chipman1992,Chipman1993,Almeida1993} that the grazing incidence reflection of X-rays on classical (perfect reflection) optics based on Wolter-I configuration, as far as described by Fresnel equations, does not introduce a spurious polarization larger than 0.1$\%$. 
Moreover, \cite{Katsuta2009} demonstrated experimentally that the reflection on multilayer (Bragg reflection) optics induces a negligible spurious polarization. They performed a reflectance measurement on a multilayer mirror sector finding no more than 0.8$\%$ of polarization. On the contrary, by using a complete mirror module (as we did with a classical mirror) an even lower spurious polarization would be expected owing to the axial symmetry that nearly (on-axis, even completely) cancel out the artificial polarization term. This consideration ensures us for future employment of the GPD, also for hard X-rays, coupled with multilayer optics. During our measurement campaign at PANTER it was not possible to verify the level of absence of spurious polarization induced by optics, because this measurement requires a rigorous control of the set-up, especially concerning the Bremsstrahlung radiation source (that is itself polarized). We are planning a new campaign aimed at addressing this specific issue.

Inhomogeneities of the optics can induce artificially a polarization, but we expect at a level negligible for our observations.
A source of spurious polarization is the non-uniformity of reflectivity. We did not perform a pencil-beam scan on the surfaces of the mirror, but no apparent asymmetry was seen in the focal spot \citep{Spiga2013}, and also the polishing process of mandrels from which the mirrors were replicated should rule out axial asymmetries in reflectivity performance. Even in the worst case, indeed, the reflectivity would change by only a few percents, and for sure it would not exhibit a bipolar distribution. Hence, any artificial polarization related to this effect should be averaged out to much less than 0.1$\%$.
Artificial polarization can derive also from profile errors of the optics in the axial direction. Such errors degrade also the angular resolution. They directly affect the incidence/reflection angles at which the X-rays strike on the mirror, so in principle they locally change the polarization of the reflected radiation. Nevertheless, the intrinsic angular resolution of the JET-X FM2 implies slope errors in the range of 10 arcsec or so. Even at the largest nominal incidence angle in the JET-X FM2 module (0.67~deg, corrected for the finite distance at PANTER), the maximum polarization introduced would be less than 0.4$\%$ in the band 0.5--10~keV. In  addition, a variation of the incidence angle, like mentioned above, locally increases the polarized term by less than 1 part over 10$^4$. The effect of roundness errors is even smaller, by a factor of 50 or more.
Therefore, to the level of sensitivity of the GPD, the expected polarizing effect of the JET-X optics is negligible ($\ll 0.5 \%$).

Another source of degradation of the polarimetric response of the GPD couple with an optics depends on scattering from roughness and from the telescope structure that may contaminate the observation of faint sources when close to bright sources. However, the fine imaging capabilities of the GPD allows for taking into account the additional contribution to the image, if any.

\section{Science goals of imaging polarimetry}\label{sec:astroobservations}

In this section we show two examples for which X-ray polarimetry resolved in space can contribute in modelling the sources, by simulating images of the Crab PWN and Cas A SNR convoluted with the PSF of the optics coupled with the GPD.
We simulated also polarimetry by using the XIPE detector configuration. Some regions of interest of the sources are taken as an example to verify the polarimetric capability of this small mission. 
The polarimetric simulation assumes the performance of the GPD as simulated by means of a Monte Carlo software, whose predictions were confirmed by experimental measurements performed in our laboratory \citep{Muleri2008}.
The uncertainties of the degree and angle of polarization of the simulated measurements are derived by convolving the response of the polarimeter with the degree of polarization expected by models and applying the Poisson statistic to the count rate of the expected signal \citep{Dovciak2011}.

\subsection{Pulsar Wind Nebulae}\label{sec:PWN}

PWNe are originated by the interaction of the relativistic particles of the pulsar wind with the interstellar medium and they are sources of non-thermal emission, ranging from radio to $\gamma$-rays, due to Synchrotron and Inverse Compton processes.
The prototype of this class of sources is the Crab Nebula for which the first positive detection of integrated polarization in the X-rays was performed in the 1971 during a sounding rocket experiment \citep{Novick1972}. 

Even from the first images of the Einstein Observatory \citep{Harnden1984} and ROSAT \citep{Hester1995} the Crab PWN appeared as a structured source, this was confirmed by the high resolution X-ray images of Chandra, that showed also the presence of small scale features \citep{Weisskopf2000a}. The complexity of PWNe structure depends on their nature. They host a neutron star surrounded by a bright axisymmetric nebulosity that comprises \textit{jets}, a structure of equatorial bright rings and a \textit{torus} \citep{Helfand2001, Gaensler2002} with wisps. These features are generated by the activity of the inner neutron star whose relativistic wind interacts with the interstellar medium, that is modelled by many MHD simulations \citep{Komissarov2003, Komissarov2004, DelZanna2004}. 
Due to the complexity of such astrophysical targets, polarimetry integrated in a wide field is no more sufficient to provide informations about all the observed features. 
PWNe are very structured in X-rays that are well suited to study the magnetic field configuration from the pulsar up to the external torus and the jets.
Imaging polarimetry, if combined with synthetic polarization maps derived from relativistic MHD simulations, would be a powerful tool for diagnostics of synchrotron emission features in PWNe \citep{Bucciantini2005}. In particular it would allow to infer the inner bulk flow structure \citep{Volpi2009} and turbulence \citep{Shibata2003}.

\begin{figure} 
 \begin{center}
\begin{tabular}{c}
\includegraphics[scale=0.5]{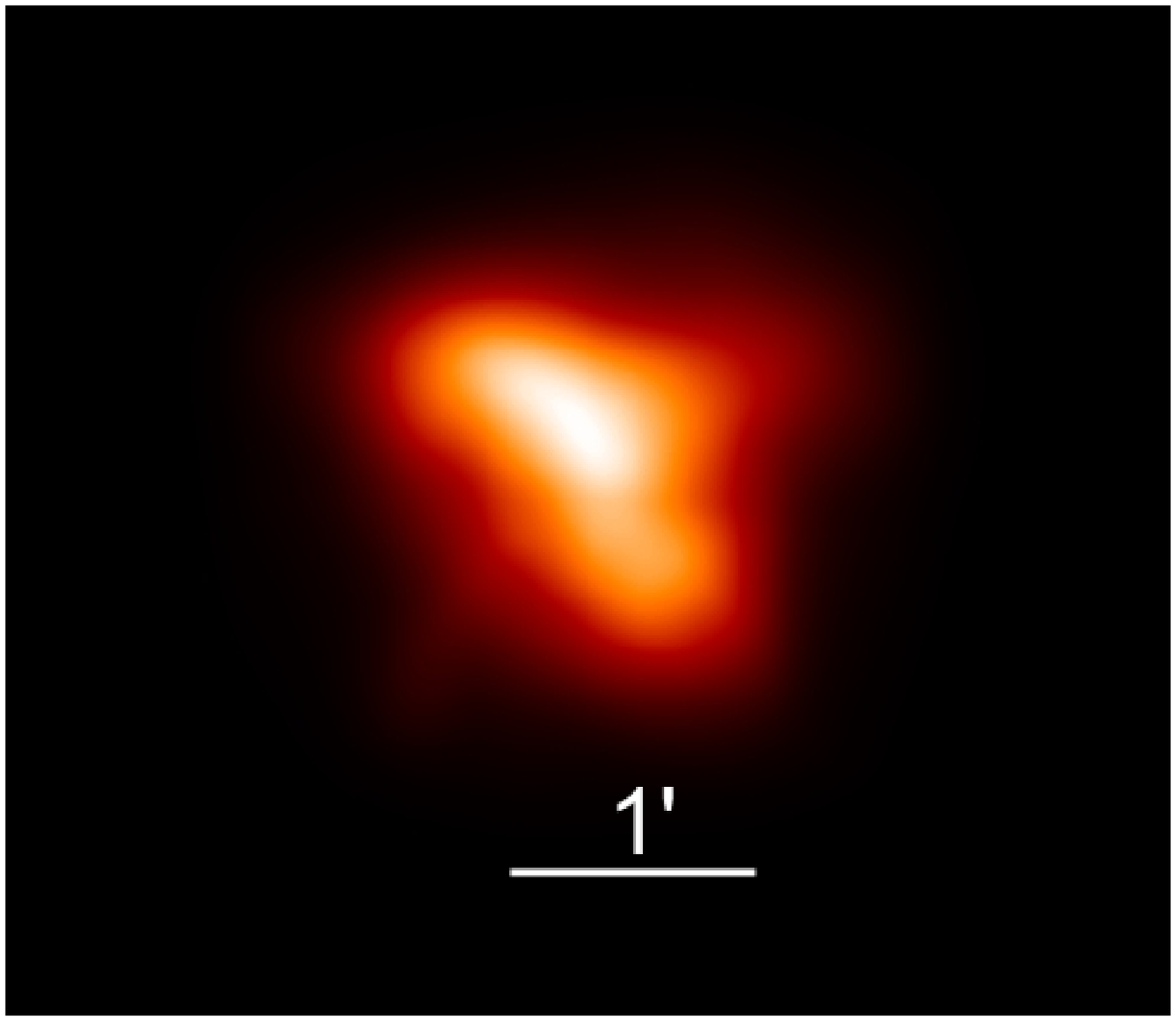} \\
\includegraphics[scale=0.24]{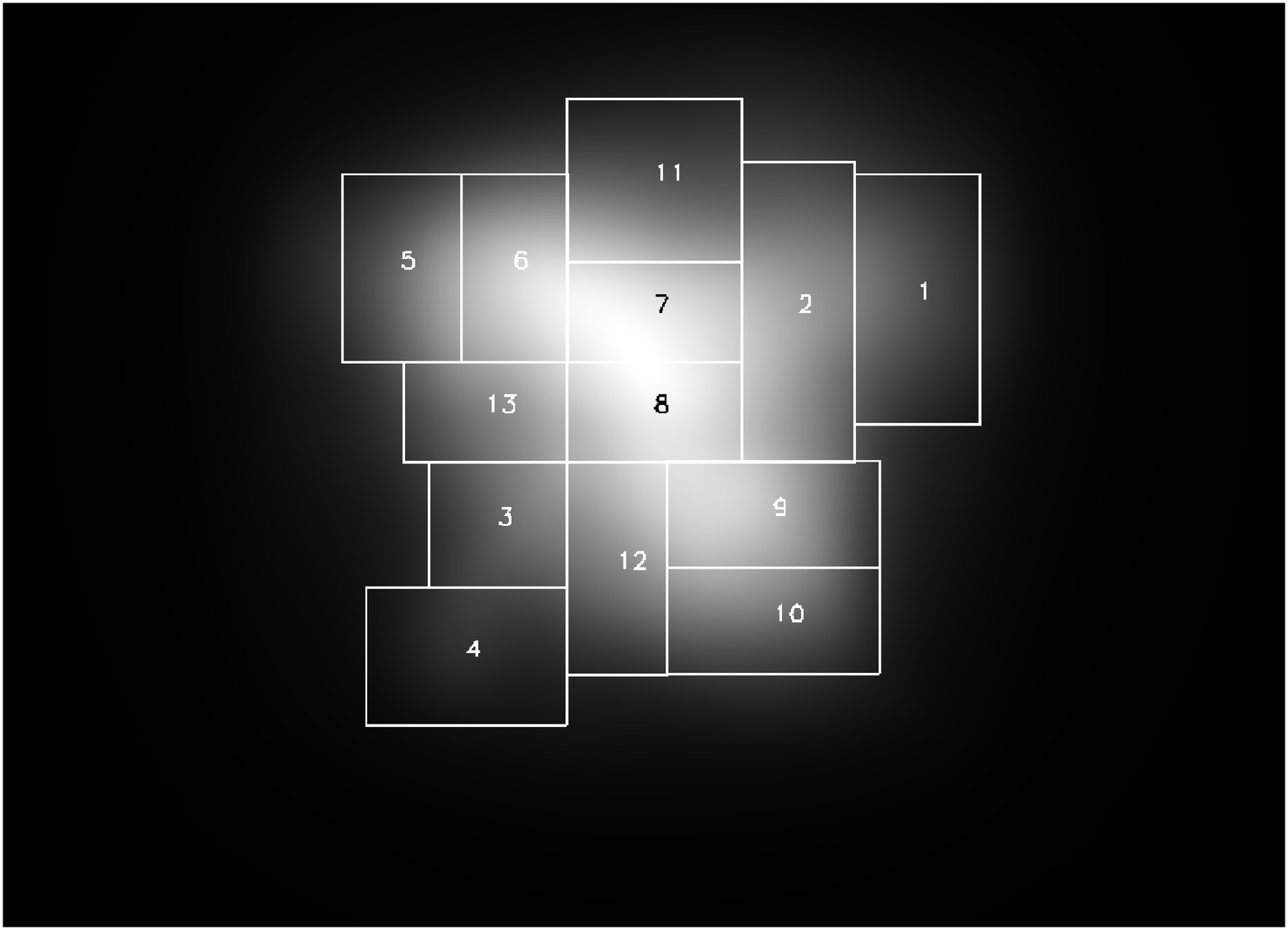} 
\end{tabular}
\caption{Top panel: The Chandra high resolution image of the Crab PWN  as it would appear at the  JET-X telescope coupled with the GPD. The image is convoluted with the on axis PSF at 4.51 keV of the XIPE polarimeter (see Fig.~\ref{fig:PSFFit} for the function profile and Tab.~\ref{tab:PSFFit} for the function parameters). The field of View of XIPE is much larger (see Tab.~\ref{tab:jetx}). The torus region is clearly separated by the jets.
Bottom panel: Selected region of the Crab image for which the polarization measurement was simulated. Results are shown in Tab.~\ref{tab:RegionCrab}
Credit: NASA/CXC/MSFC/M.Weisskopf et al \label{fig:Crab}}
 \end{center}
 \end{figure}

The GPD at the focus of a suitable telescope could address these issues. In fact, two detectors were proposed at the focal plane of two JET-X optics modules, in the framework of the XIPE mission proposal \citep{Soffitta2013b} to the ESA Call of 2012 for a launch of a small mission in 2017. 
In Fig.~\ref{fig:Crab} (top panel) we blurred the Chandra image of the Crab PWN with the PSF of XIPE. The original high resolution image is convoluted with the on axis PSF at 4.51 keV of the XIPE polarimeter that is shown in Fig.~\ref{fig:PSFFit} (see Tab.~\ref{tab:PSFFit} for the PSF parameters). 
Even if the smaller features are not resolved, the torus region is separated by the jets and the polarization across the image can be studied for different regions as shown in Fig.~\ref{fig:Crab} (bottom panel).
We simulated an observation by assuming a polarization degree of $19\%$ \citep{Weisskopf1978} in the energy range 2-10 keV for a 100 ks observation. 
We chose the average polarization measured by \cite{Weisskopf1978} to be conservative.
The blurred image of the source has been subdivided in 13 regions and for each one the $1-\sigma$ errors of the degree and angle of polarization have been evaluated by simulation and reported in Tab.~\ref{tab:RegionCrab}, together with the corresponding MDP (Minimum Detectable Polarization at 99$\%$ of confidence level, see \cite{Weisskopf2010}) for the measurement.

We refer to the synthetic polarization maps of PWNe by \cite{Volpi2009} derived by means of relativistic MHD simulations.
They assumed a purely toroidal magnetic field of the torus, thus the polarization vector is parallel to its symmetry axis and rotate by increasing the angular distance from this axis up to become
orthogonal.
The angular position at which the $90^\circ$ rotation is completed is a function of the pulsar wind flow velocity, for a given inclination angle of the torus.
For lower bulk flow velocity, this turning point is further away from the symmetry axis.
Such a swing of the polarization vector could be detected easily by XIPE by looking at regions No. 5, 6, 7, 8, 9 and 10 of Fig.~\ref{fig:Crab} (bottom panel).
Polarization is expected to be very high (up to 70$\%$) at the center (regions 7 and 8) and to decrease moving to the edges (regions 5 and 10, respectively). The higher is the bulk velocity, the sharper is the polarization decrease. 
The net effect of such a blurring is to sum the signals coming from slightly different positions and this obviously implies a certain depolarization if, as expected from current models, the polarization is different for each feature of the source. 
On the one hand, the angle of polarization should change smoothly passing from a region to another and major rotations should occur on angular scale larger than our PSF. 
Moreover, the high degree of polarization measured with OSO-8 \citep{Weisskopf1978} by averaging the signal from the entire source points to the fact that, if a rotation of the angle of polarization does exist, it is not very large, or, alternatively, the polarization degree from some bright region is much higher than that from the others so that a net polarization "survives". Therefore, we expect that the contamination of the polarization signal from neighbour regions does not affect significantly our sensitivity estimates reported in Tab.~\ref{tab:RegionCrab}.

Imaging capability of the GPD is, moreover, a powerful tool for the measurement of the polarization of the pulsar emission alone. The pulsed signal from the inner neutron star (corresponding to a pulsed fraction of about 15$\%$ at 2.6 keV and 11$\%$ at 5.2 keV, \cite{Weisskopf1978}) can be studied by means of a phase-resolved analysis and, thanks to imaging, the contribution of the nebular emission can be reduced by selecting only the region where the pulsar is located. The nebular emission acts like a background and by reducing its contribution with respect to the pulsar signal, it is possible to improve the sensitivity of the measurement. \cite{Elsner2012} addressed this issue by performing a detailed statistical analysis and considering the specific case of the Crab pulsar.

As seen in Sect.~\ref{sec:onaxis}~and~\ref{sec:offaxis}, the image quality depends largely on the telescope intrinsic PSF. Therefore, future availability of telescopes with a better angular resolution (few arcseconds), and possibly large effective area, would allow a more detailed study of extended sources such as PWNe. For example, with IXO/ATHENA-like optics (5~arcsec of intrinsic angular resolution, \cite{Collon2011, Ghigo2012}), small features such as the bright knots would be observable singularly and detailed polarization maps could be obtained. For IXO, due to the long focal length (20 m) and very narrow penetration angles, the blurring for inclined absorption in gas would be negligible and the GPD would preserve totally the telescope performance in terms of image resolution \citep{Lazzarotto2010}. A small worsening is expected for 
ATHENA+\footnote{\url{$\mathrm{http://athena2.irap.omp.eu/IMG/pdf/sp_14.pdf}$}} due to the larger penetration angles that are consequence of the shorter focal length (12 m) and the optimization for a softer energy band. However, as demonstrated for JET-X, this worsening gives an angular resolution only few arcseconds larger that the one of the telescope, up to grazing incidence angle of the order of the degree. Therefore, it is reasonable to expect the total angular resolution to be not worst than 7~arcsec--8~arcsec.

At the present 54 PWNe are known and many presents interesting features like the Crab PWN \citep{Kargaltsev2008}.
Among them Jellyﬁsh, Vela, G0.9+0.1/G0.87+0.08 and MSH 11–62/G291.02–0.11 (No. 7, 13, 41 and 48 in \cite{Kargaltsev2008}) are mCrab sources having a size comparable to the Crab PWN. 
For these sources an uncertainty on the polarization angle of about 10$^\circ$ (instead of about 1$^\circ$ of the Crab) could be reached if observed for 500 ks each and if divided in the half of the number of regions with respect to the Crab example (to double the flux for each region).
This is due to the fact that such an uncertainty scales as the root square of the number of counts \citep{Weisskopf2010, Strohmayer2013}, therefore of the integration time (or of intensity, or of the effective area).
This measurement was never performed so far and is compatible with the science objective of a pathfinder mission like XIPE.

Other sources like Kes 75, Mouse and G54.10+0.27 (No. 12, 22, 10 in \cite{Kargaltsev2008}) are also mCrab sources, but their extension is less then about 1/3 of the Crab. Therefore, they can be easily studied by means of IXO/ATHENA-like optics having an effective area 100 times larger then XIPE and an angular resolution more than 3 times better. Moreover, many other faint but extended sources, down to about 0.01 mCrab, would be accessible with this advanced optics (for example some of them are G21.50–0.89, 3C 58, G106.65+2.96, G332.50–0.28, G11.18–0.35 and IC 443/G189.23+2.90 corresponding to No.4, 5, 6, 8, 14 and 47  in \cite{Kargaltsev2008}).

 \begin{deluxetable}{crrc}
\tabletypesize{\scriptsize}
\tablecaption{Simulation of a polarization measurement for the Crab. The source is subdivided in 13 regions as shown in Fig.~\ref{fig:Crab} (bottom panel). The uncertainties of the degree and angle of polarization are listed, assuming a polarization degree of $19\%$ \citep{Weisskopf1978} in the energy range 2-10 keV for a 100 ks observation. \label{tab:RegionCrab}}
\tablewidth{0pt}
\tablehead{\colhead{Region No.} & \colhead{$\sigma_{\mathrm{degree}}$ ($\%$)} & \colhead{$\sigma_{\mathrm{angle}}$ (deg)} & \colhead{ MDP ($\%$)} }
\startdata
\hline
1 & 0.7 & 1.1 & 2.2 \\ 
2 & 0.5 & 0.8 & 1.5 \\ 
3 & 0.8 & 1.3 & 2.5 \\ 
4 & 1.0 & 1.6 & 3.2 \\ 
5 & 0.7 & 1.1 & 2.2 \\ 
6 & 0.5 & 0.9 & 1.7 \\
7 & 0.5 & 0.8 & 1.6 \\  
8 & 0.5 & 0.8 & 1.6 \\ 
9 & 0.5 & 0.9 & 1.7 \\
10 & 0.7 & 1.1 & 2.2 \\ 
11 & 0.6 & 1.0 & 1.9 \\
12 & 0.6 & 1.0 & 1.9 \\
13 & 0.7 & 1.1 & 2.2 \\    
\enddata
\end{deluxetable}

\subsection{Shell-like Supernova Remnants}

In shell-like SNRs environment electrons are accelerated up to 10-100 TeV by shocks and they radiate via synchrotron emission up to X-rays \citep{Reynolds1981}.
Therefore, polarization maps would give information about the particles acceleration processes and the magnetic field behaviour in this turbulent environment \citep{Bykov2009}. 
Above 10 keV the thermal component (bremsstrahlung and lines) is superseded by a power-law extending up hundreds of keV \citep{Vink2005, Helder2008} that originates hard tails in shell-like SNRs (including Cas A) due, possibly, to synchrotron emission \citep{Allen1997, Allen1999a}.
However, the non-thermal fraction of the spectrum can be highlighted by means of polarimetry also at lower energy, if regions with a minor line emission are identified in the images. In these cases it would be possible to study the non-thermal component not only by means of spectral analysis.

\begin{figure} 
 \begin{center}
\begin{tabular}{c}
\includegraphics[scale=0.55]{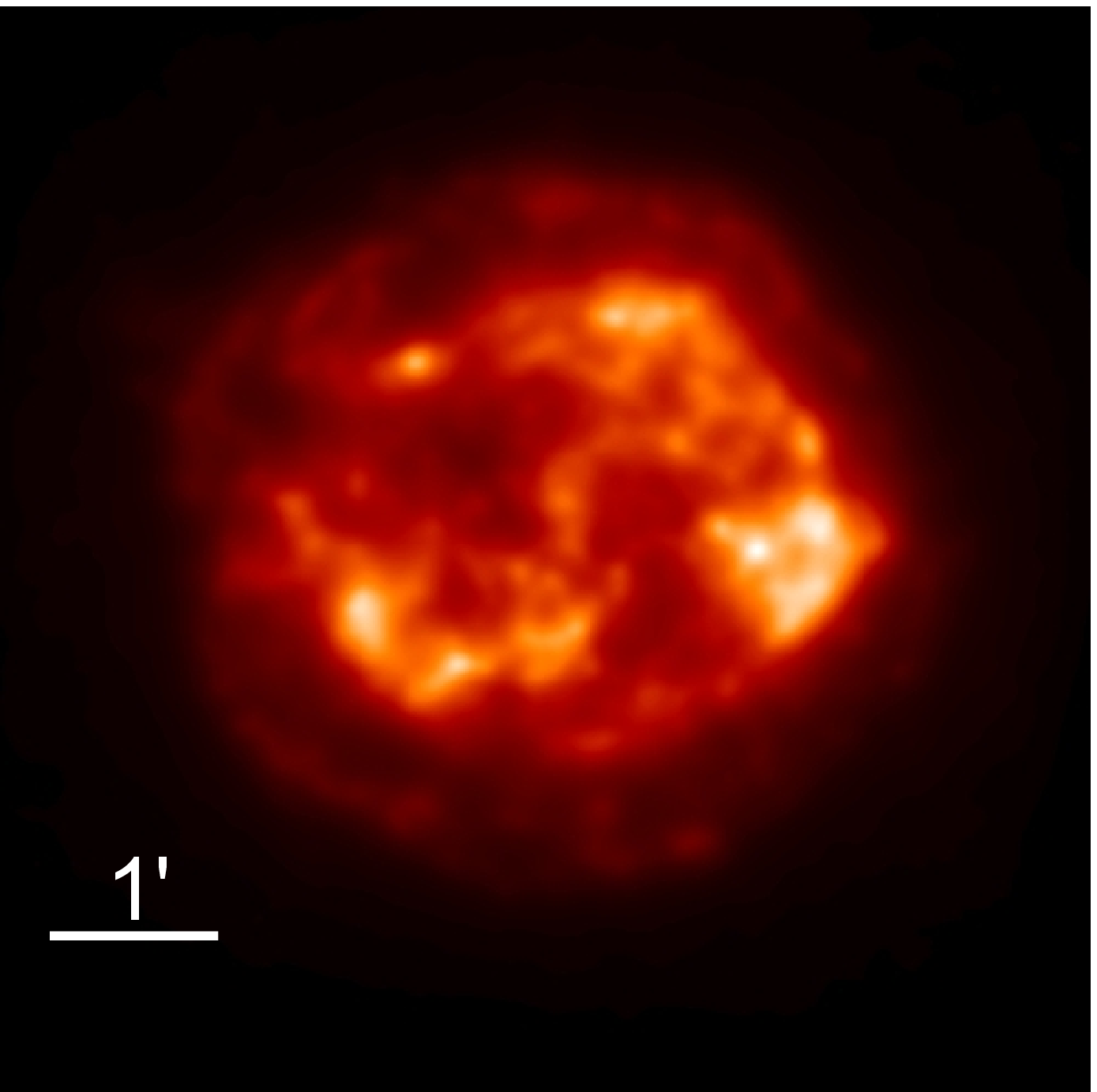}\\
\includegraphics[scale=0.215]{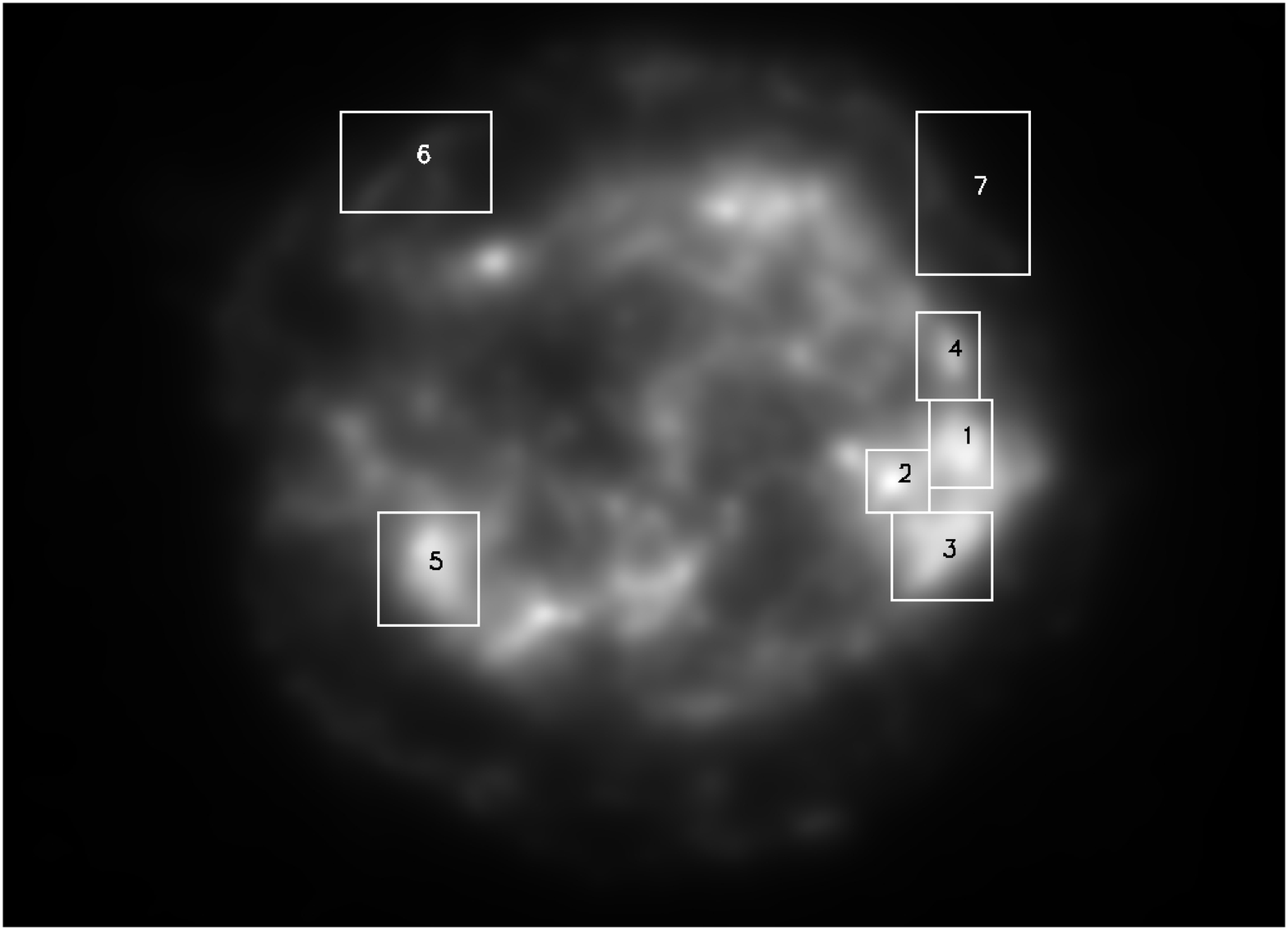}
\end{tabular}
\caption{Top panel: The Chandra high resolution 4--6 keV image of the Cassiopeia A SNR  as it would appear at the JET-X telescope coupled with the GPD. The image is convoluted with the on axis PSF at 4.51 keV of the XIPE polarimeter (see Fig.~\ref{fig:PSFFit} for the function profile and Tab.~\ref{tab:PSFFit} for the function parameters). The SNR is clearly resolved and its features can be studied separately. The entire source fits in the field of view of the polarimeter (see Tab.~\ref{tab:jetx}).
Bottom panel: Selected region of the Cassiopeia A SNR for which the polarization measurement was simulated. Results are shown in Tab.~\ref{tab:RegionCasA}
Credit: NASA/CXC/MSFC/M.Weisskopf et al \label{fig:CasA}}
 \end{center}
 \end{figure}

The polarized non-thermal emission would come primarily from structures such as filaments and clumps and the capability to perform imaging polarimetry would allow to resolve the emission coming from different structural features.
Particularly suited for this kind of study are the young SNRs like Cas-A, Tycho and Kepler that efficiently accelerate electrons and show the X-ray emissions from well confined narrow regions \citep{Araya2010a}. 
Such non-thermal emitting regions were observed by Chandra at the edges of the Cas-A shell-like SNR. In the images they are particularly evident in the 4--6 keV continuum dominated band \citep{Vink2005} and spatially coincide with the radio emitting region \citep{Gotthelf2001}. The absence of line emission in this spectral region is an evidence for the synchrotron origin of the radiation \citep{Vink2003}. Similar features were found also in other shell-like young SNRs such as SN 1006, Tycho and Kepler \citep{Bamba2003, Hwang2002, Cassam2004}.
In Fig.~\ref{fig:CasA} (top panel) we blurred the Chandra image of Cas A (4--6 keV) with the PSF of XIPE. The image is convoluted with the on axis PSF at 4.51 keV as done for the Crab PWN. This SNR is clearly resolved and its features can be studied separately.
The entire source fits in the field of view of XIPE (see Tab.~\ref{tab:jetx}), therefore the polarimetric measurement can be performed with a single observation.
\cite{Bykov2009} simulated a synthetic model of SNR, deriving the expected polarization maps at different energies for synchrotron emission. The turbulent nature of the SNR environment produce granular features in the magnetic field configuration that give rise to similar features in the polarization map. In the model of \cite{Bykov2009} polarization degree can reach high values (up to 50$\%$) in correspondence of some of them.
To evaluate the polarized fraction of the total emission (thermal plus non-thermal) that is effectively observed, we must consider that the non-thermal component was evaluated to be the 29$\%$ of the thermal Bremsstrahlung flux of the entire source in the 4--6 keV energy band by \cite{Bleeker2001} (see also \cite{Helder2008}) and that it corresponds to the 22.5$\%$ of the total flux. Therefore, by assuming 50$\%$ of polarization for the non-thermal component in a bright emitting feature, a polarization degree of about 11$\%$ would arise from the total emission.
This value was used to simulate a measurement for a 2 Ms observation of Cas-A. We selected 7 interesting regions of the source image (see Fig.~\ref{fig:CasA}, bottom panel). Some of them are bright spots, whose emission is given by the over-position of both thermal and non-thermal components (regions No. 1, 2, 3 and 5 on the "disc" of the shell-like SNR). 
The other regions numbered as 4, 6 and 7 have spectra with low or negligible line emission (see \cite{Hughes2000} for region 4 and \cite{Araya2010a} for regions 6 and 7), therefore they are probably dominated by the non-thermal component. In these cases the thermal component fraction could be smaller than previously evaluated and the polarization arising from their emission should be higher. 
The resulting $1-\sigma$ errors of the degree and angle of polarization are reported in Tab.~\ref{tab:RegionCasA} with the corresponding MDP.

 \begin{deluxetable}{crrc}
\tabletypesize{\scriptsize}
\tablecaption{Simulation of a polarization measurement for Cas-A. The source is subdivided in 7 regions as shown in Fig.~\ref{fig:CasA} (bottom panel). The uncertainties of the degree and angle of polarization are listed, assuming a polarization degree of 11$\%$ in the energy range 4-6 keV for a 2 Ms observation.
Regions 4, 6 and 7 are probably dominated by the non-thermal component, therefore the polarization arising from their emission should be higher with respect to regions 1,2,3 and 5 in which the thermal component is dominant.
 \label{tab:RegionCasA}}
\tablewidth{0pt}
\tablehead{\colhead{Region No.} & \colhead{$\sigma_{\mathrm{degree}}$ ($\%$)} & \colhead{$\sigma_{\mathrm{angle}}$ (deg)} & \colhead{ MDP ($\%$)} }
\startdata
\hline
1 & 2.4 & 6.6 & 7.7 \\ 
2 & 2.7 & 8.3 & 8.8 \\ 
3 & 2.1 & 5.9 & 6.7\\ 
4 & 2.9 & 7.8 & 9.5 \\ 
5 & 1.9 & 5.3 & 6.1 \\ 
6 &3.5 & 11.0 & 11.1  \\
7 & 3.6 &11.0 & 11.6 \\  
\enddata
\end{deluxetable}

The JET-X optics coupled with the GPD are capable to obtain a good angular resolution suited to study extended SNR such as Cas A, that fits entirely in the field of view (see Tab.~\ref{tab:jetx}) and therefore is observable with a single pointing. A further improvement would require optics modules with a larger effective area, that if associated with a better angular resolution, would allow to study the fine structure of bright emitting features or to access to fainter filaments ignored in this analysis. 
While in the case of a bright source such as the Crab Nebula the angular resolution limits, at the present, a finer imaging polarimetry study, in the case of fainter extended source such as SNRs, the effective area of the optics is the driving parameter for further improvements. However, large missions are usually designed with combined instruments for different kind of astrophysical target and the time dedicated to polarimetry is usually far smaller than that of a dedicated mission like XIPE. 
In any case, since polarimetry needs a high count statistics, the effective area requirement is a priority with respect to the angular resolution, in fact, a count statistics good to do the image, could not be sufficient for polarimetry.

\section{Conclusions}

The GPD is a photoelectric polarimeter with an intrinsic imaging capability that makes it suitable to be used as a focal plane instrument to do the image of the source while performing polarimetry in the X-rays. 
So far, the imaging performance of the GPD coupled with X-ray telescopes were only studied by means of Montecarlo simulations and for the first time we demonstrated its feasibility by means of experimental measurements. 

We measured the Point Spread Function of the GPD placed at the focus of the JET-X X-ray telescope in the 2-10 keV energy band, both for on-axis and off-axis radiation beams. 
This detector/optics system is the configuration proposed for the pathfinder mission XIPE as response to the ESA small mission Call of 2012.
We measured the angular resolution in terms of HEW, that is 22.7~arcsec at 2.98 keV, 23.2~arcsec at 4.51 keV and 28.9~arcsec at 8.05 keV for on axis radiation. 
In this work we showed that a detector/optics configuration typical of a pathfinder mission is able to obtain important results, opening the field of imaging polarimetry also in the X-rays. PWNe and SNRs were considered as case studies and the relation between the polarized emission, the source geometry and the magnetic field configuration were analysed.

We demonstrated experimentally that the image quality of the optical system given by the GPD coupled to an X-ray mirror module depends mainly on the telescope intrinsic PSF. We showed that even with a small mission like XIPE bright SNRs and PWNe can be studied by imaging polarimetry.
The availability of optics with a better angular resolution (few arcseconds) and a large effective area would allow GPD to obtain even more detailed images while performing sensitive polarimetry for many extended sources. Therefore, with IXO/ATHENA-like optics smaller and fainter features of a larger population of sources would be accessible.

\acknowledgments
\section*{Acknowledgements}
S. Fabiani and F. Muleri acknowledge the STSM (Short Term Scientific Mission) program of the COST (European Cooperation in Science and Technology) Action MP1104: ``Polarization as a tool to study the solar system and beyond" which financially supported the measurement campaign at the PANTER facility.

\bibliographystyle{apj}
\bibliography{References}

\clearpage

\end{document}